\documentclass[sigconf,screen,nonacm]{acmart}
\usepackage{subcaption}
\usepackage{enumitem}
\usepackage[most]{tcolorbox}
\usepackage{acronym}
\usepackage{balance}
\usepackage{hyperref}
\usepackage{xurl}
\usepackage[utf8]{inputenc}
\hypersetup{
    breaklinks=true % URL 줄바꿈 시 링크가 깨지지 않고 이어지도록 설정
}

\newif\ifhighlight
\highlightfalse

\usepackage{xcolor}
\ifhighlight
  \newcommand{\revised}[1]{{\color{blue}#1}}
  \newcommand{\deleted}[2][]{{\color{red}\sout{#2}}} % 필요시 ulem 패키지 추가
\else
  \newcommand{\revised}[1]{#1}
  \newcommand{\deleted}[2][]{} % 삭제된 내용은 표시 안 함
\fi

\AtBeginDocument{%
  }

\begin{document}

\title{CLEAR: Causal Context-Based Agentic Reasoning for Vulnerability Detection}
\titlenote{This research was supported by the MSIT(Ministry of Science and ICT), Korea, under the Convergence security core talent training business support program(IITP-2024-RS-2024-00423071, 10\%) supervised by the IITP(Institute of Information \& Communications Technology Planning \& Evaluation), This work was supported by the National Research Foundation of Korea(NRF) grant funded by the Korea government(MSIT)(RS-2024-00356926, 50\%), and This work was supported by the Institute of Information \& Communications Technology Planning \& Evaluation(IITP) grant funded by the Korea government(MSIT) (RS-2024-00341722, Development of Cyber Resilience Method for Intelligent Service Robots, 40\%).}

\author{Sungju Yun}
\authornote{Both authors contributed equally to this research.}
\orcid{0009-0000-8752-0129}
\affiliation{%
    \institution{Hanyang University}
    % \department{Department of Computer Science \& Engineering}
    % Major in Bio Artificial Intelligence
    \city{Ansan}
    % postal 15588
    \country{Republic of Korea}
}
\email{tjdwn77777@hanyang.ac.kr}

\author{Sijune Hwang}
\authornotemark[2]
\orcid{0009-0006-0059-0305}
\affiliation{%
    \institution{Hanyang University}
    % \department{Department of Information Security}
    \city{Ansan}
    % postal 15588
    \country{Republic of Korea}
}
\email{ghkdtlwns987@hanyang.ac.kr}

\author{Yeonjoon Lee}
% \correspondingauthor
\authornote{Corresponding author.}
\orcid{0000-0002-5010-8277}
\affiliation{%
    \institution{Hanyang University}
    % \department{College of Computing}
    \city{Ansan}
    % postal 15588
    \country{Republic of Korea}
}
\email{yeonjoon@hanyang.ac.kr}

\author{Kyungtae Kang}
\orcid{0000-0002-6587-7044}
\affiliation{%
    \institution{Hanyang University}
    % \department{Department of Artificial Intelligence}
    \city{Ansan}
    % postal 15588
    \country{Republic of Korea}
}
\email{ktkang@hanyang.ac.kr}

\author{Sungbin Park}
\orcid{0009-0000-8140-8822}
\affiliation{
    \institution{Hanyang University}
    % \department{Department of Computer Science \& Engineering}
    % Major in Bio Artificial Intelligence
    \city{Ansan}
    % postal 15588
    \country{Republic of Korea}
}
\email{pbt98@hanyang.ac.kr}

\begin{abstract}
Detecting source code vulnerabilities is increasingly difficult as modern security flaws are rooted in complex causal dependencies between execution flows, control conditions, and program states. Despite recent advances in Large Language Models (LLMs) and multi-agent frameworks, existing approaches primarily address superficial similarities between benign and vulnerable functions while failing to capture the complex causal dependencies inherent in security flaws. To address these limitations, we propose \textbf{Causal Context-based Agentic Reasoning (CLEAR)}, a novel multi-agent vulnerability detection framework integrated with a causal knowledge graph. CLEAR systematically constructs a \textbf{Vulnerability Causal Knowledge Graph (VCKG)} that models the causal chains between entrypoints, preconditions, root causes, and fix intents across vulnerability instances. Leveraging this structured knowledge, four specialized agents including the Collector, Claim, Critic, and Judge collaboratively verify vulnerability hypotheses through retrieved causal contexts. Experimental results on C/C++ and Java vulnerability benchmarks demonstrate that CLEAR improves Pair-Correct (P-C) performance by 130.7\% and 71.56\% over state-of-the-art approaches, respectively, demonstrating the effectiveness of causal knowledge graph–guided reasoning for automated vulnerability detection.
\end{abstract}

%%
%% The code below is generated by the tool at http://dl.acm.org/ccs.cfm.
%% Please copy and paste the code instead of the example below.
%%


% \ccsdesc[500]{Do Not Use This Code~Generate the Correct Terms for Your Paper}
% \ccsdesc[300]{Do Not Use This Code~Generate the Correct Terms for Your Paper}
% \ccsdesc{Do Not Use This Code~Generate the Correct Terms for Your Paper}
% \ccsdesc[100]{Do Not Use This Code~Generate the Correct Terms for Your Paper}

%%
%% Keywords. The author(s) should pick words that accurately describe
%% the work being presented. Separate the keywords with commas.

\maketitle

\section{Introduction}

As modern software systems grow in scale and complexity, accurately identifying and classifying vulnerabilities within source code has become a fundamental task for ensuring system security. While manual inspection by security professionals provides high-quality results, it is difficult to scale across massive codebases due to the significant time and human resources required \cite{johnson2013don}. Moreover, the increasing complexity of vulnerabilities makes it challenging to maintain consistent detection accuracy through manual review alone. In response, various automation studies have been conducted, ranging from traditional static and dynamic analysis \cite{goseva2015capability, kaur2020comparative} to deep learning-based methods \cite{li2018vuldeepecker, zou2021interpreting, chakraborty2021deep, Ni2023DistinguishingLI, steenhoek2023empirical, rahman2024towards, risse2024uncovering}. More recently, with the remarkable success of Large Language Models (LLMs) in various NLP tasks \cite{lewis2020retrieval, zheng2023survey, neumann2024llm, llmstalk, wan2025digest}, researchers have increasingly explored their application to automated vulnerability detection \cite{tamberg2025harnessing, llms_cannot_reason_vuln, llm_future_direction, sheng2025llms}.

Despite these advances, accurately distinguishing vulnerable code from its non-vulnerable counterparts remains a non-trivial challenge. In practice, many software vulnerabilities in real-world scenarios arise from subtle differences between benign and vulnerable functions, where the core flaw is often hidden behind nearly identical syntactic structures\cite{tan2024}. Recent benchmarks such as PrimeVul\cite{primevul} have been specifically curated to reflect this reality by providing datasets where the functional differences between benign and vulnerable code pairs are extremely subtle. However, existing models often exhibit significant performance degradation when evaluated on such paired instances because they struggle to capture these fine-grained distinctions and fail to precisely differentiate a secure implementation from its flawed counterpart. This performance gap in challenging benchmark settings suggests that robust vulnerability detection requires a deep analysis of the structural causal nuances that define a security flaw rather than relying solely on high-level semantic understanding.

To address this issue, recent frameworks such as VulTrial \cite{widyasari2025let} have introduced multi-agent debate mechanisms that emulate courtroom style reasoning, aiming to mitigate overconfidence, hallucination, and the lack of verification observed in single-agent approaches \cite{improving, liang-etal-2024-encouraging, he2025llm, wan-etal-2025-mamm}. By enabling agents to generate competing hypotheses and critically examine each other’s claims, collaborative argumentation helps surface semantic inconsistencies between similar code variants. However, these multi-agent approaches still fundamentally rely on the internal pre-trained knowledge of LLMs, making them inherently susceptible to reasoning degeneration and bias \cite{huang2023large, huang2025survey}. While they leverage the code understanding capabilities of LLMs, they lack an explicit mechanism to resolve the complex causal dependencies that actually define a vulnerability, often leading to false positives when encountering subtle code similarities.

\vspace{2mm}

\begin{figure*}[t]
  \centering
  \includegraphics[width=\linewidth]{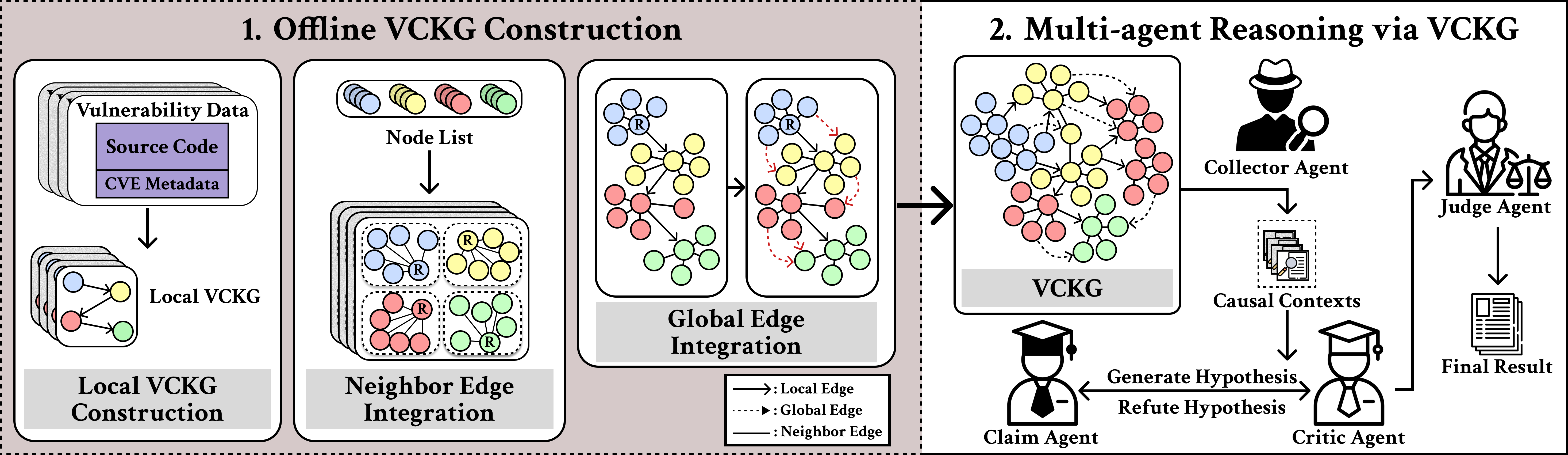}
  \Description{An Overview of the CLEAR Framework}
  \caption{An Overview of the CLEAR Framework}
  \label{fig:CLEAR overview}
\end{figure*}

To overcome these limitations, we propose \textbf{CLEAR (Causal Context-based Agentic Reasoning)}, a framework designed to perform vulnerability detection by grounding multi-agent reasoning in structured causal contexts. CLEAR moves beyond simple semantic pattern matching by utilizing the underlying causal dependencies such as the relationships between entrypoints, preconditions, and root causes to resolve the ambiguity between similar code variants. To support this process, we introduce the \textbf{Vulnerability Causal Knowledge Graph (VCKG)}, which systematically models the causal chains of security flaws across diverse vulnerability instances. By representing vulnerabilities as structured causal paths rather than fragmented code snippets, VCKG provides an external knowledge base that constrains the reasoning process, allowing LLMs to focus on the logical mechanisms of an exploit rather than superficial code similarities.

We conduct comprehensive evaluations against both single-agent and multi-agent baselines. For single-agent methods, we consider deep learning–based models such as CodeBERT \cite{feng2020codebert} and UniXCoder \cite{guo2022unixcoder}, as well as LLM-based approaches following the prompting strategy of Ding et al. \cite{primevul}. For multi-agent frameworks, we benchmark against GPTLens \cite{hu2023large} and the state-of-the-art model VulTrial \cite{widyasari2025let}. Evaluation on C/C++ and curated Java-Pair datasets shows that CLEAR outperforms state-of-the-art models by 130.7\% and 71.56\%, respectively, in the Pair-Correct (P-C) metric. These results empirically validate that structural causal reasoning is essential for capturing the complex nature of security vulnerabilities and effectively distinguishing them from benign code. The primary contributions of this paper are as follows:

\vspace{5pt}

% \begin{enumerate}[leftmargin=20pt]
\begin{itemize}[
    leftmargin=20pt,
    topsep=2pt,
%     itemsep=2pt,
%     parsep=0pt,
    partopsep=0pt
]
    \item \textbf{Design of the CLEAR Framework:} We develop a causal knowledge-based multi-agent analysis framework that utilizes the causal context retrieved from VCKG as a reasoning constraint to perform multi-perspective vulnerability verification.
    \item \textbf{VCKG Modeling:} We propose the first Vulnerability Causal Knowledge Graph (VCKG), which organizes causal relationships among related vulnerability types into a structured format optimized for LLM reasoning.
    \item \textbf{Empirical Validation:} We demonstrate the effectiveness of structural causal reasoning by achieving up to a 130\%  performance improvement over existing SOTA models across diverse programming language datasets, including C/C++ and Java.
\end{itemize}
% [Section 2.0] R1 C2 VCKG Design Clarification: CLEAR is fundamentally capable of implicitly modeling these nuances at the attribute level. For instance, "improper input handling" maps directly to a Precondition node (missing environmental safe states), while "sanitization loopholes (corner cases)" are structured within the descriptive text details of the Root Cause node. We acknowledge this capability was not sufficiently emphasized in our text, and we commit to clarifying these mapping definitions within Section 2.0 (Design of Vulnerability Causal Knowledge Graph).

% [Section 2.1] R3 Q1. Train/Test Separation: We deeply appreciate the reviewer highlighting this critical point. The VCKG is constructed solely from the training split. Test code slices are strictly unseen queries and never enter the graph. We will clarify this split policy in Section 2.1.

% [Section 2.1 Local VCKG Construction] R1 C1. Prompt for Node Extraction: We appreciate this remark. The exact prompts and system instructions utilized by the LLM for VCKG entity and edge extraction will be explicitly integrated into the Appendix.

\vspace{-0.5em}
\section{Methodology}

In this section, we present the overall architecture of CLEAR, a framework designed to perform vulnerability detection through multi-agent reasoning guided by structured causal contexts retrieved from a Vulnerability Causal Knowledge Graph (VCKG). The framework consists of two primary processes: (1) VCKG Construction and (2) Multi-Agent Reasoning. In the VCKG construction phase, the system operates in an offline environment, taking raw vulnerability data as input to map the complex causal chains and propagation paths of vulnerability data. Subsequently, during the Multi-Agent Reasoning phase, CLEAR leverages the structured causal contexts retrieved from the VCKG to support hypothesis generation, adversarial verification, and final decision making through collaborative interaction among specialized agents, enabling vulnerability detection.

\vspace{-5pt}
% =================================================================
\setcounter{subsection}{-1}
\subsection{Design of Vulnerability Causal Knowledge Graph}
% =================================================================

Vulnerability detection fundamentally requires understanding the causal chain from an external trigger to a logical failure. However, existing LLM-based approaches often rely on internally learned statistical patterns and may therefore struggle to explicitly capture the underlying causal structure of vulnerability formation \cite{rahman2024towards, risse2024uncovering}. Incorporating structured context through retrieval can support LLM reasoning by grounding the inference process in explicitly retrieved vulnerability-related contexts \cite{gao2023retrieval, ovadia2024fine}, thereby reducing reliance on implicit statistical associations. However, conventional retrieval strategies typically rely on flat semantic similarity or chunk-level matching, which may introduce irrelevant context or fail to preserve the essential causal relationships required for reliable vulnerability reasoning \cite{wang2025causalrag}.

To address these challenges, we propose the Vulnerability Causal Knowledge Graph (VCKG), which organizes vulnerability data into structured causal components, including execution entrypoints, triggering pre conditions, root causes, and fix intents. This representation explicitly captures causal dependencies involved in vulnerability formation and provides structured context to support reasoning in retrieval-based LLM frameworks. \revised{At a granular level, VCKG implicitly models fine-grained vulnerability characteristics at the attribute level by systematic mapping: for instance, operational constraints such as "improper input handling" map directly to a Precondition node as a missing environmental safe state, while specific functional flaws like "sanitization loopholes (corner cases)" are structured within the descriptive text details of the corresponding Root Cause node.}

\begin{figure}[htbp]
    \centering
    \vspace{-8pt}
    \includegraphics[width=0.6\linewidth]{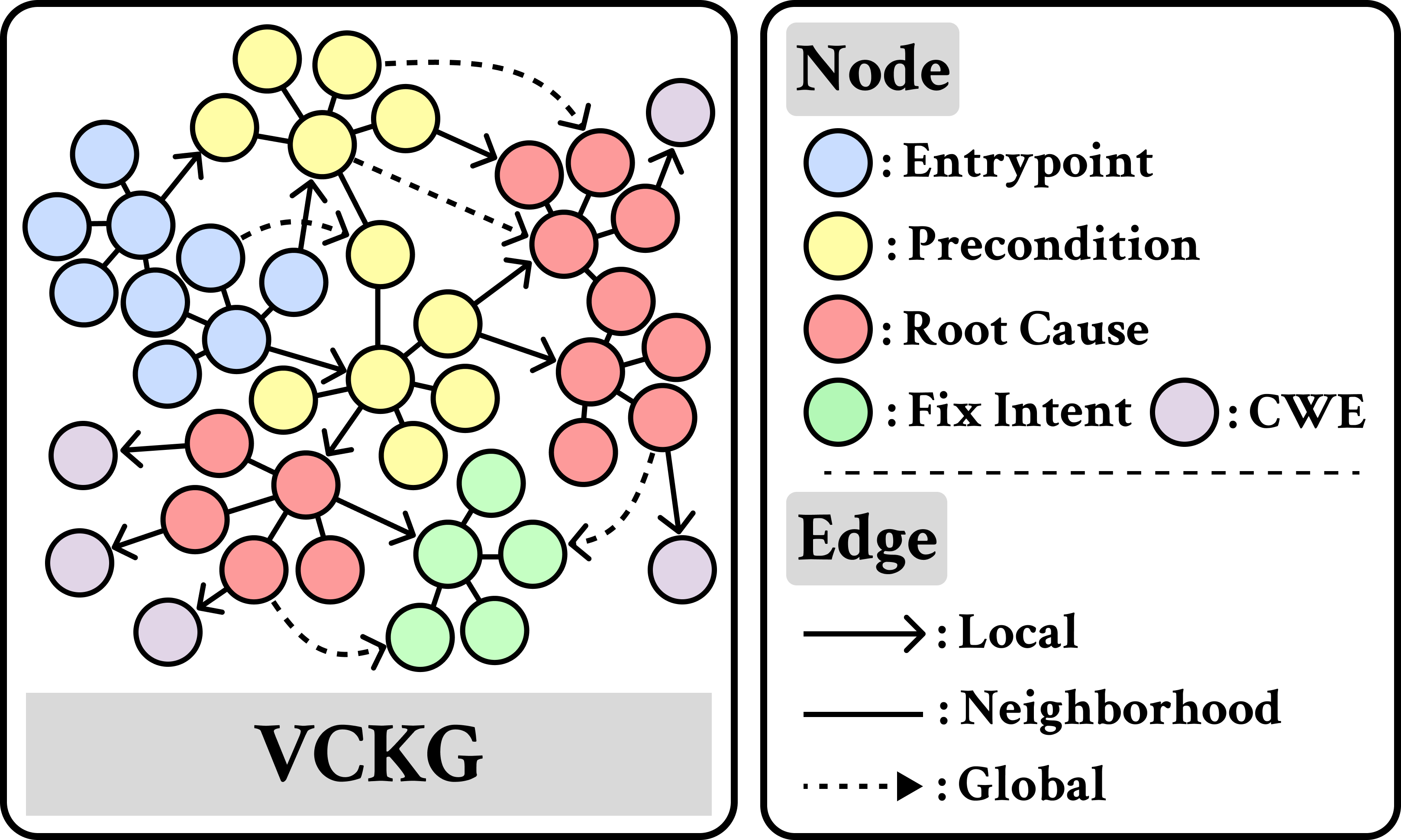}
    \Description{Design of Vulnerability Causal Knowledge Graph}
    \caption{Design of Vulnerability Causal Knowledge Graph}
    \label{fig:Design of VCKG}
\end{figure}

\textbf{Causal Entities (Nodes).} To capture the causal structural nature of security flaws, VCKG defines five specialized nodes. These entities represent the essential components of a vulnerability's lifecycle:
\begin{itemize}[leftmargin=20pt]
    \item \textbf{Entrypoint (EP):} The initial execution context or interface where untrusted input enters the system.
    \item \textbf{Precondition (PC):} The specific environmental states or program conditions required for a flaw to be reachable or triggered.
    \item \textbf{Root Cause (RC):} The core logical error or architectural weakness that inherently constitutes the security flaw.
    \item \textbf{Fix Intent (FI):} The developer’s underlying logic and the targeted patch logic applied to neutralize the flaw, serving as the primary differentiator between vulnerable and safe code.
    \item \textbf{CWE:} A standardized abstraction that provides high-level semantic categorization for the specific weakness.
\end{itemize}

\textbf{Causal Relations (Edges)} 
The interactions between these entities are defined by three distinct edge types, which categorize the nature of the causal and semantic links:

\begin{itemize}[leftmargin=20pt]
    \item \textbf{Local Edge:} Represents the internal causal flow within a single vulnerability instance (e.g., within the same function or CVE case). It connects $EP \rightarrow PC \rightarrow RC \rightarrow FI$ and $CWE$ to map the direct propagation path and its corresponding remedy.

    \item \textbf{Neighborhood Edge:} Connects causal entities based on shared semantic properties or logical proximity (excluding CWE). This edge facilitates the discovery of cross-component associations that are not captured by direct execution flows.

    \item \textbf{Global Edge:} Establishes relationships between different Local VCKG instances. It links similar entities across the knowledge base, allowing the model to leverage external causal patterns and previous fix strategies for the current analysis.
\end{itemize}

The design of VCKG is centered on the insight that vulnerabilities are context-dependent causal events. By explicitly modeling the Fix Intent (FI) and its local causal relationship with the Root Cause (RC), VCKG provides the LLM with the logical "reason" why a patch works. This prevents the model from misidentifying a patched version as vulnerable simply because it shares a high percentage of code tokens with the original version. Furthermore, the categorization of edges into Local, Global, and Neighborhood ensures a balance between deep internal analysis and broad contextual reasoning, making the detection process logically rigorous rather than merely statistical.

\begin{figure}[htbp]
  \centering
  \includegraphics[width=\linewidth]{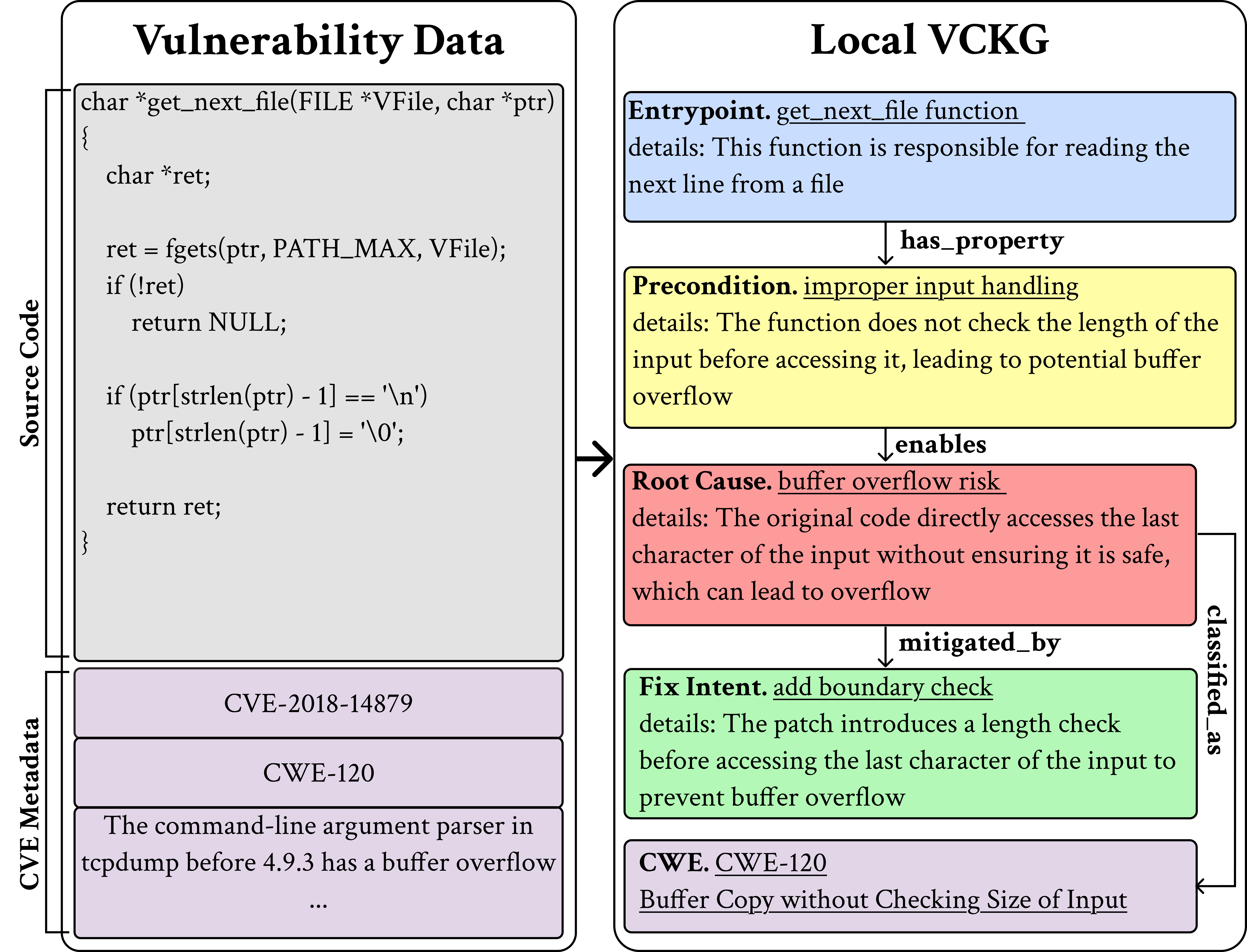}
  \Description{Local Vulnerability Causal Knowledge Graph Example}
  \caption{Local Vulnerability Causal Knowledge Graph Example}
  \label{fig:Local VCKG Example}
\end{figure}

\vspace{-10pt}

% =================================================================
\subsection{Offline Vulnerability Causal Knowledge Graph Construction}
% =================================================================

The construction of the VCKG is performed in an offline environment to ensure a robust and comprehensive knowledge base before the reasoning process begins. This phase transforms raw vulnerability data—consisting of source code, CVE descriptions, and patch information—into a structured causal network. The construction process is divided into three sequential stages: (1) Local VCKG Construction, (2) Neighbor Edge Integration, and (3) Global Edge Integration. \revised{To prevent data leakage and guarantee evaluation integrity, the VCKG is constructed solely from the training split; test code slices are treated strictly as unseen queries and are completely excluded from the graph construction phase.}

% =================================================================
\textbf{Local VCKG Construction.} The first stage focuses on extracting individual causal chains from specific vulnerability instances, where each Local VCKG corresponds to a single vulnerability instance represented by a function-level code snippet aligned with its associated CVE metadata and patch information. In this setting, one Local VCKG is constructed per vulnerability-containing function-level code snippet, ensuring that each snippet contributes exactly one isolated causal chain representation. Using this vulnerability data as input, a Large Language Model (LLM) analyzes the logical dependencies within the source code and descriptive metadata to identify the five core causal entities (EP, PC, RC, FI, and CWE) and determine their roles within the vulnerability lifecycle. \revised{Due to paper page constraints, the complete prompt templates and system instructions utilized for local VCKG construction are provided in the Supplementary Material.}

As illustrated in Figure \ref{fig:Local VCKG Example}, this stage maps the internal propagation path by establishing Local Edges ($EP \rightarrow PC \rightarrow RC \rightarrow FI \text{ and } CWE$), resulting in a set of independent "Local VCKG" units. Each unit represents the isolated causal logic of a single security flaw, providing a granular foundation for the subsequent integration phases.

% =================================================================
\textbf{Neighborhood Edge Integration.}
% =================================================================
While individual Local VCKGs capture the internal causal context of specific vulnerabilities, establishing inter-graph connectivity is essential for constructing a robust and high-performing knowledge base. Neighborhood Edge Integration is the process of linking semantically related causal nodes (EP, PC, RC, FI) across different Local VCKGs to form a cohesive network. To preserve causal-role consistency during graph expansion, Neighborhood Edges are constructed only between nodes of the same causal type. Restricting alignment to same-type nodes prevents structurally incompatible cross-role associations and improves the precision and stability of node-wise causal-context retrieval in the subsequent multi-agent reasoning process. However, performing a pairwise semantic similarity comparison across all nodes in the entire knowledge base is computationally expensive and impractical for large-scale datasets. To ensure both efficiency and precision, we implement two strategic constraints:

\vspace{-0.5em}

\begin{itemize}[leftmargin=20pt]
    \item \textbf{CWE-based Clustering:} We first organize Local VCKGs into clusters based on their associated CWE Map as defined by the MITRE framework. Semantic similarity comparisons are then restricted to causal nodes within the same CWE cluster, significantly reducing the search space while ensuring that only logically related vulnerabilities are compared.
    
    \item \textbf{Embedding-based Top-$k$ Filtering:} Within each cluster, we compute semantic embeddings for the nodes and establish Neighbor Edges only for the Top-$k$ most similar counterparts. The value of $k=5$ was selected to maintain a compact causal neighborhood while preventing excessive context growth that may otherwise introduce redundant information and reduce the effectiveness of downstream LLM-based reasoning.
\end{itemize}

Through this integration, semantically similar nodes are interconnected into a ``neighborhood`` structure. This allows the framework to retrieve not only the target causal chain but also its most relevant neighbors, providing the reasoning agents with a broader spectrum of similar attack patterns and fix strategies during the subsequent detection phase.

\vspace{2mm}

\begin{figure}[t]
  \centering
  \includegraphics[width=0.8\linewidth]{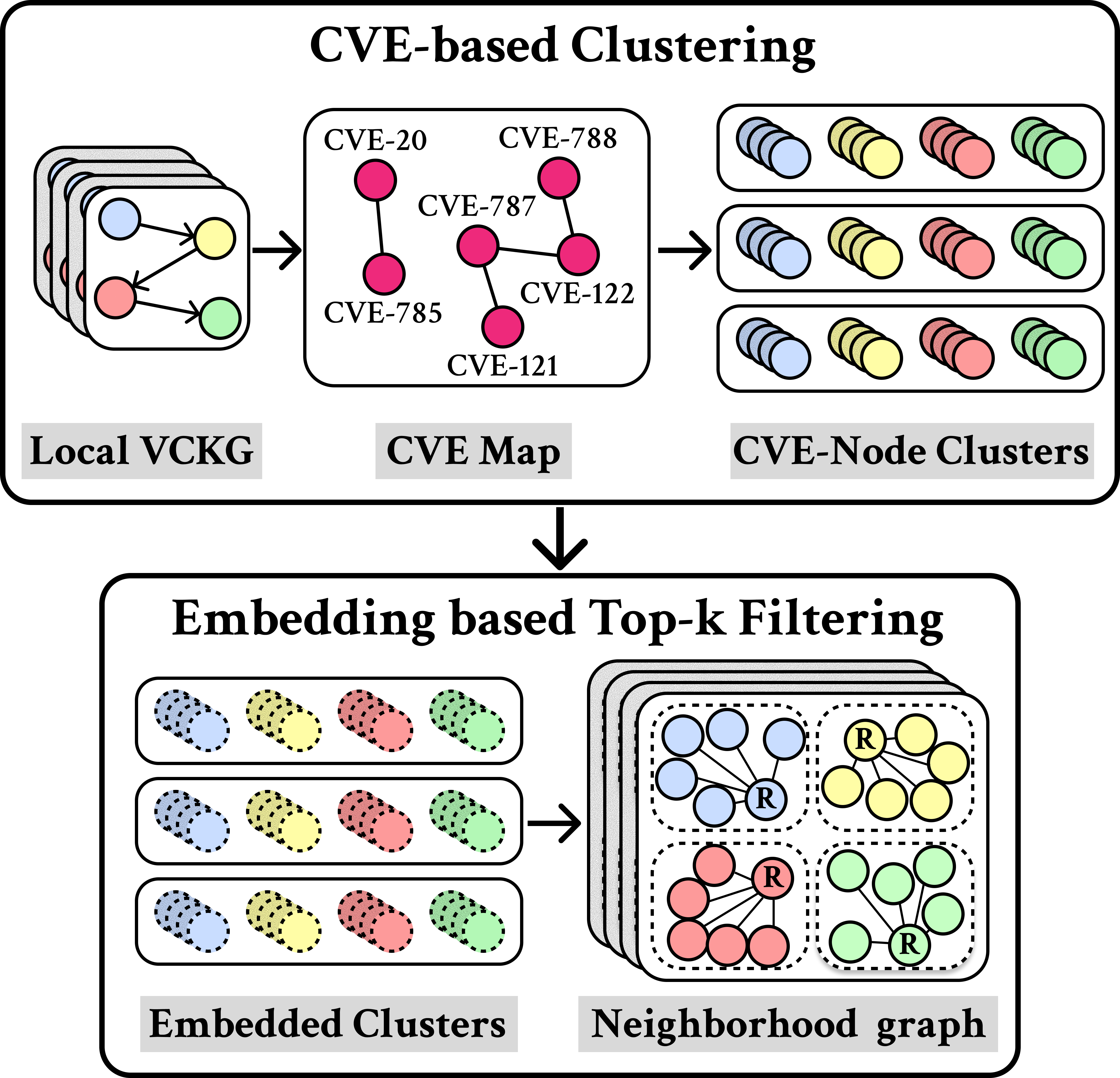}
  \Description{Neighborhood Edge Integration}
  \caption{Neighborhood Edge Integration}
  \label{fig:Neighborhood Edge Integration}
\end{figure}

% \vspace{2mm}

\begin{figure}[htbp]
  \centering
  \vspace{-5pt}
  \includegraphics[width=\linewidth]{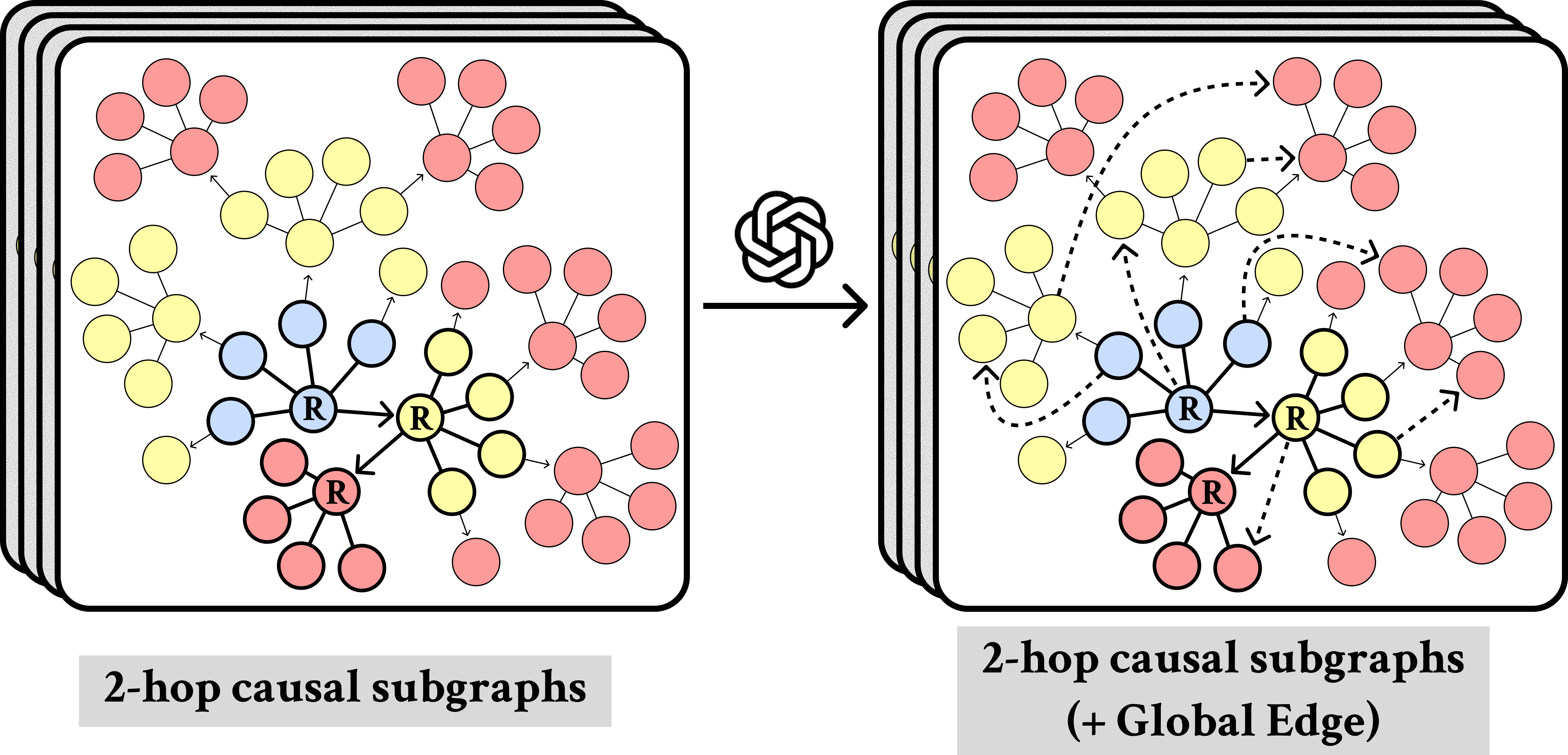}
  \Description{Global Edge Integration}
  \caption{Global Edge Integration}
  \label{fig:Global Edge Integration}
\end{figure}

% =================================================================
\textbf{Global Edge Integration.}  The final stage of VCKG construction is Global Edge Integration, a process driven by an LLM to establish causal links between nodes in different Local VCKGs. While Local Edges map the internal causal chains within a single vulnerability data, Global Edges extend these same causal dependencies across the entire knowledge base by linking entities from disparate Local VCKG units that exhibit functional or logical inter-dependencies.

To ensure computational feasibility and manage the LLM's context window efficiency, we perform this integration by querying the LLM with 2-hop causal subgraphs as the fundamental unit of input. Rather than processing the entire global network at once, the system extracts localized 2-hop paths—such as $Entrypoint \rightarrow Precondition \rightarrow Root\ Cause$ or $Precondition \rightarrow Root\ Cause \rightarrow Fix\ Intent$, including their respective semantic neighborhoods. This windowing approach provides a concentrated context of approximately 260 nodes, allowing the LLM to effectively identify latent causal overlaps where an entity in one instance functionally influences a node in another.

Through this process, Vulnerability Data is transformed from a collection of isolated, unstructured text into a structurally well-defined Knowledge Graph that inherently possesses deep causal insights. As illustrated in Figure \ref{fig:Global Edge Integration}, this 2-hop strategy prevents the exponential expansion of the search space while maintaining a scalable foundation for the subsequent multi-agent reasoning process. By synthesizing these cross-instance links, the VCKG evolves from a set of fragmented cases into a highly interconnected causal network.

% =================================================================
\subsection{Multi-Agent Reasoning via VCKG}
% =================================================================
The Multi-Agent Reasoning phase is the core analytical process where a specialized team of agents, including the Collector, Claim, Critic, and Judge, collaborates to identify vulnerabilities by leveraging structured causal insights from the VCKG. As illustrated in Figure~\ref{fig:CLEAR overview}, this phase is initiated by the Collector Agent, which performs a deep Causal Context Retrieval to provide the reasoning agents with highly relevant historical evidence grounded in the target code's logic. 

\begin{figure}[htbp]
  \centering
  \includegraphics[width=0.9\linewidth]{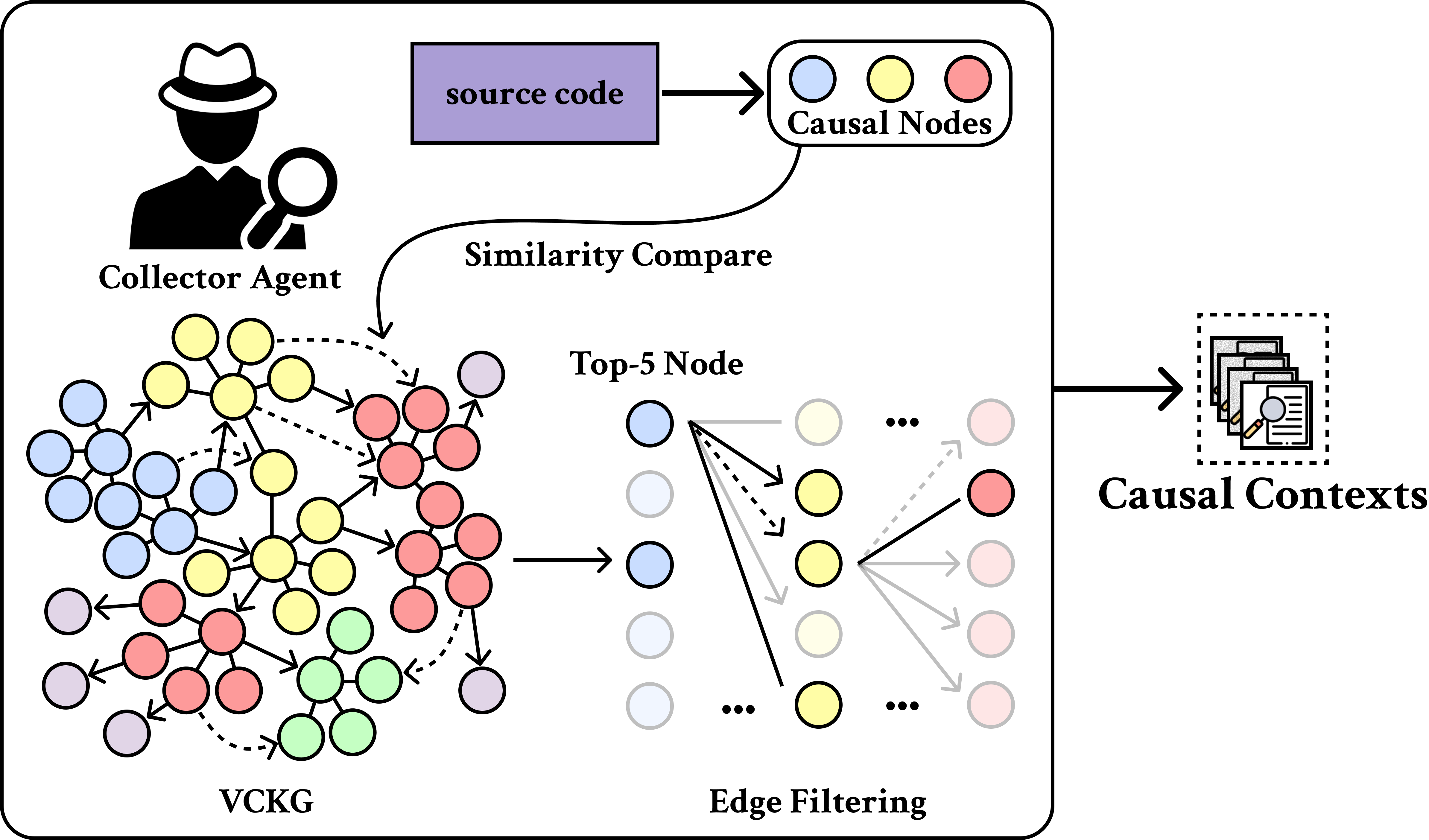}
  \Description{Causal Context Retrieval}
  \caption{Causal Context Retrieval}
  \label{fig:Causal Context Retrieval}
\end{figure}

As illustrated in Figure~\ref{fig:Causal Context Retrieval} the workflow begins with the Collector analyzing the target source code to extract Causal Nodes, with a specific focus on Entrypoints (EP), using the same LLM-driven reasoning mechanism described in the construction phase. Once the local EP nodes are identified, the Collector interacts with the global VCKG to perform a semantic similarity comparison, retrieving the Top-5 most similar historical EP nodes. These retrieved nodes serve as anchors to traverse all associated Local, Neighbor, and Global edges, gathering a comprehensive set of next-hop causal entities.

\vspace{-0.4em}

To ensure the precision of the reasoning process, the Collector employs an LLM-driven Edge Filtering strategy. Instead of delivering all retrieved nodes, the Collector queries the LLM to evaluate whether each next-hop entity maintains a genuine logical relevance to the specific context of the current source code. Any node deemed irrelevant through this query-based judgement is filtered out.

If this filtering process results in all candidate nodes being removed, the Collector executes a secondary retrieval phase as a fallback mechanism. In this stage, the system re-initiates the similarity comparison using other extracted causal nodes, specifically Precondition (PC) and Root Cause (RC), to identify their respective Top-5 most similar historical counterparts within the VCKG. This iterative expansion ensures that the subsequent agents are provided with validated Causal Contexts even when the initial entrypoint-based search is overly restrictive.

Following this robust retrieval process, the Claim Agent utilizes the validated Causal Contexts as logical anchors to generate a specific vulnerability hypothesis. Unlike conventional prompting that relies on the LLM’s internal parameters, the Claim Agent is constrained to map the target code's execution flow onto the retrieved causal paths (e.g., $EP \rightarrow PC \rightarrow RC$), ensuring the hypothesis is grounded in verifiable historical patterns.

To mitigate potential hallucinations or overconfidence, the Critic Agent then performs an adversarial verification. It attempts to refute the proposed hypothesis by identifying logical inconsistencies between the target code's program state and the retrieved causal preconditions, specifically searching for missing fix intents (FI) or unmet preconditions that would invalidate the exploit.

Finally, the Judge Agent reviews the arguments from both sides along with the supporting causal context to reach a final decision. By weighing the Critic’s objections against the Claim’s evidence, the Judge determines whether a vulnerability truly exists, identifies its specific type (CWE), and explains the underlying cause. This collaborative process ensures that the final result is not just a statistical guess but a conclusion backed by a clear and verifiable reasoning chain.

\vspace{2mm}

\begin{figure*}[t]
  \centering
  \includegraphics[width=\linewidth]{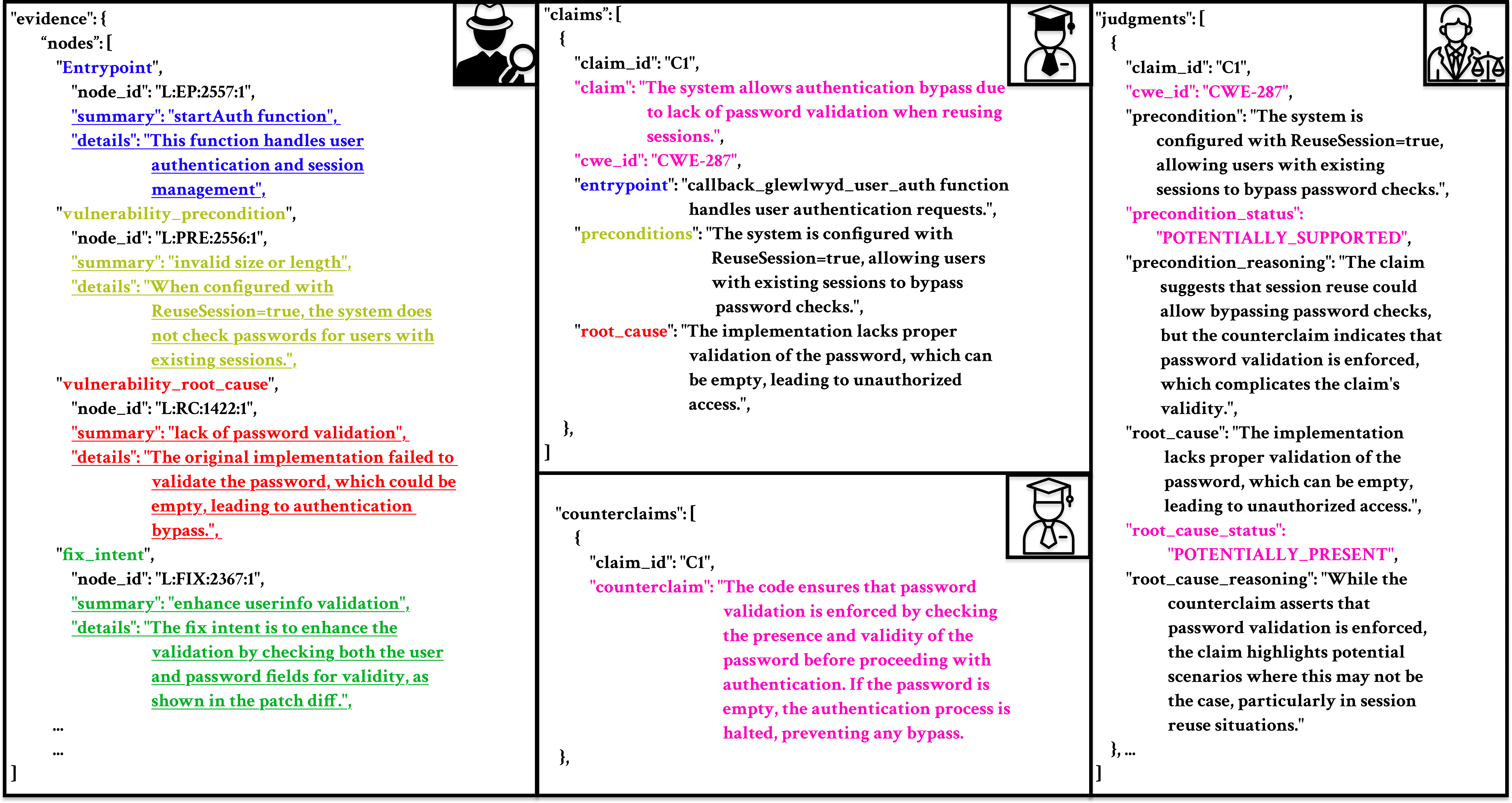}
  \Description{Multi-Agent Reasoning Example}
  \caption{Multi-Agent Reasoning Example}
  \label{fig:Multi-Agent Reasoning Example}
\end{figure*} 
% [3.2 Datasets] R3 C1. Small Java Test Set & Sampling Noise: To eliminate sampling bias, we will execute 10 repeated runs for the Java-Pair evaluation and report variance, confidence intervals, and standard deviations.

% [3.4 Baseline Methods] VulRAG 추가에 따른 변화

\vspace{-0.4em}

\section{Experimental Setup}
This section describes the experimental configuration used to evaluate the effectiveness of CLEAR, including the implementation environment, datasets, evaluation metrics, and baseline methods.

\vspace{-0.4em}
\subsection{Implementation Details}
All implementations were conducted in \texttt{Python 3.12}. We used \texttt{LangChain}~\cite{LangChain} to construct the multi-agent reasoning pipeline. To ensure fair comparisons across all experimental settings, including baselines and ablation variants, all experiments were performed using the same backbone model, \texttt{gpt-4o-mini}, with \texttt{temperature=0.0}. For embedding-based retrieval and knowledge graph construction, we used \texttt{all-MiniLM-L6-v2}, a Sentence-BERT–based embedding model~\cite{reimers2019sentence}. The same embedding model was consistently applied to the Collector Agent, vulnerability causal knowledge graph construction, and RAG-based baseline comparisons to maintain experimental consistency.

\vspace{-0.4em}
\subsection{Datasets}
For the C/C++ setting, we evaluated our approach on the PrimeVul dataset~\cite{primevul}, which contains 7,578 training samples, 960 validation samples, and 870 test samples. Following the official pairwise evaluation protocol of PrimeVul, vulnerability detection was evaluated using 433 vulnerable/non-vulnerable pairs constructed from the test split (435 vulnerable and 435 non-vulnerable samples). For CWE classification, evaluation was performed on the vulnerable subset of the test split, consisting of 435 vulnerable samples. 

For the Java setting, we constructed a curated Java-Pair dataset by integrating CVEFixes~\cite{cvefixes}, MegaVul~\cite{ni2024megavul}, and CleanVul~\cite{li2024cleanvul}. Since CleanVul does not provide CVE or CWE identifiers, we enriched vulnerability metadata by matching commit identifiers using the NVD API~\cite{nist_nvd_api} and OSV API~\cite{osv_database}. To focus on subtle logic-level vulnerability differences, vulnerable/non-vulnerable pairs were filtered using string similarity $\geq 0.8$, resulting in 214 training samples, 27 validation samples, and 132 test samples. Vulnerability detection was evaluated using 66 vulnerable/non-vulnerable pairs constructed from the test split, while CWE classification was evaluated on the 66 vulnerable samples. \revised{Given the relatively small sample size of the Java-Pair dataset, a single evaluation run could be highly susceptible to statistical volatility. To mitigate this and ensure reliable generalization, we repeated the evaluation 10 times over independent iterations and reported the metrics with their corresponding confidence intervals.}

\vspace{-0.4em}
\subsection{Evaluation Metrics}
Following the pairwise evaluation perspective of PrimeVul \cite{primevul}, we report separate metrics for vulnerability detection and CWE classification.

\textbf{Detection Metrics.} We use pairwise metrics to reflect whether the model correctly distinguishes between vulnerable and non-vulnerable code pairs. Pairwise Correct Prediction (P-C) measures fully correct pair decisions, while P-V and P-B capture degenerate behaviors where both snippets are predicted as vulnerable or non-vulnerable, respectively. Pairwise Reversed Prediction (P-R) penalizes failure cases in which the vulnerable/non-vulnerable roles are flipped. We additionally report Vulnerability Pair Score (VPS), which summarizes useful discrimination by rewarding correct pair ordering and discounting reversed predictions. All pairwise metrics are reported as percentages (\%).

\textbf{Classification Metrics.} For CWE classification, class imbalance is substantial, and performance must be assessed from multiple angles. Macro-F1 treats all CWE classes equally and highlights robustness on minority classes. Weighted-F1 reflects overall practical performance by accounting for class support. Matthews Correlation Coefficient (MCC) is included as a balanced correlation-based metric that remains informative under skewed label distributions and complements F1-based analysis. For consistency with the detection metrics, all reported values are expressed as percentages (\%).

\vspace{-0.4em}
\subsection{Baseline Methods} We compare three groups of baselines: \revised {Deep learning methods, single-agent methods, and multi-agent methods,}. For all baselines, we follow each paper's replication package and preserve the original prompts and evaluation logic for fair comparison. To ensure a controlled evaluation setting, all methods are executed using the same backbone LLM (\texttt{gpt-4o-mini}).

\textbf{Deep learning baselines.}
\revised{For deep learning-based baselines, we evaluate CodeBERT \cite{feng2020codebert} and UniXcoder \cite{guo2022unixcoder} to assess capabilities of deep learning models. First, we adopt CodeBERT, a bimodal pre-trained Transformer optimized via masked language modeling and replaced token detection to capture NL-PL semantic connections \cite{feng2020codebert}. We fine-tune CodeBERT on our dataset and leverage the pooled \texttt{[CLS]} token representations for pairwise discrimination. Second, we integrate UniXcoder, a unified cross-modal model that controls context access via mask attention matrices with prefix adapters, while capturing syntax and structure through bimodal sequences of mapped ASTs and code comments \cite{guo2022unixcoder}. We fine-tune UniXcoder's encoder and utilize the cosine similarity of the final-layer pooled hidden states to determine vulnerability status. Both models are implemented following their official configurations with identical input lengths and training schedules to ensure a fair comparison.}

\textbf{Single-Agent baselines.}
\revised{For single-agent baselines, we evaluate both prompting-based and Retrieval Augmented Generation based LLM approaches. For prompting-based evaluation, we adopt a Chain-of-Thought (CoT) \cite{primevul} reasoning strategy where the model directly generates a predicted CWE label step-by-step from the raw input without external structured evidence, evaluating only vulnerable samples following standard classification protocols. To assess single-agent RAG capabilities, we integrate Vul-RAG \cite{vul-rag}, a knowledge-level framework that distills high-level, generalizable vulnerability knowledge—encompassing functional semantics, root causes, and fixing solutions—from historical CVE instances. During online inference, Vul-RAG extracts semantics from the target code to retrieve relevant abstraction profiles and sequentially prompts the LLM to verify whether the target snippet exhibits the same vulnerability cause while lacking the corresponding defensive solution.}

\textbf{Multi-Agent baselines.}
For multi-agent baselines, we include GPTLens \cite{hu2023large} and VulTrial \cite{widyasari2025let}, which represent collaborative reasoning frameworks for vulnerability analysis. GPTLens performs iterative cross-agent discussion to refine vulnerability judgments through perspective diversification, while VulTrial adopts a courtroom-style debate mechanism that alternates between claim generation and counterargument verification. In our experiments, both frameworks are implemented following the configurations provided in their official replication packages. We preserve their original agent interaction protocols and prompting strategies. 

For the CWE classification task, we further follow their classification output processing pipelines to obtain a single predicted CWE label per vulnerable instance. Specifically, GPTLens produces ranked CWE candidates with confidence scores, from which we select the final prediction based on the highest ranking score according to its original evaluation procedure. Similarly, VulTrial generates candidate CWE lists through multi-agent debate outputs recorded in its classification results, which we normalize into single-label predictions by selecting the ground-truth CWE when present among the candidates, and otherwise choosing the top-ranked candidate following its replication setup.

\section{Evaluation}
\textbf{Research Question}
In this section, we evaluate the effectiveness of CLEAR through a series of experiments designed to answer the following research questions.
% \begin{itemize}[leftmargin=15pt]
\begin{itemize}[
    leftmargin=15pt,
    topsep=2pt,
    itemsep=1pt,
    parsep=0pt,
    partopsep=0pt
]
\revised{
\item \textbf{RQ1:} Does CLEAR improve vulnerability detection performance compared to existing SOTA approaches?
\item \textbf{RQ2:} How does multi-agent contribute to CLEAR's performance compared to conventional retrieval-based approaches?
\item \textbf{RQ3:} Which components of CLEAR contribute most to the performance improvement?
\item \textbf{RQ4:} How does CLEAR perform across different LLM backbones?
\item \textbf{RQ5:} How effectively can CLEAR classify vulnerability types?
\item \textbf{RQ6:} Does CLEAR construct a high-quality and reliable causal knowledge graph?
}
\end{itemize}

\vspace{2mm}

\begin{table*}[h]
\centering
\small
\caption{Pairwise vulnerability detection results across PrimeVul and Java-Pair dataset. Higher is better for P-C and VPS; lower is better for P-V, P-B, and P-R.}
\label{tab:Table1-Vulnerability-Detection}
\begin{tabular}{l l r r r r r l}
\hline
\textbf{Agents} & \textbf{Method} & \textbf{P-C ($\uparrow$)} & \textbf{P-V ($\downarrow$)} & \textbf{P-B ($\downarrow$)} & \textbf{P-R ($\downarrow$)} & \textbf{VPS ($\uparrow$)} & \textbf{Dataset} \\
\hline
DL & CodeBERT & 3.00 & 54.96 & 39.72 & 2.30 & 0.69 & PrimeVul \\
DL & UniXCoder & 6.23 & 64.66 & 28.40 & \textbf{0.69} & 5.54 & PrimeVul \\
Single-Agent & CoT & 6.47 & 67.90 & 18.94 & 6.70 & -0.23 & PrimeVul \\
\revised{Single-Agent} & \revised{Vul-RAG} & \revised{8.78} & \revised{42.26} & \revised{38.57} & \revised{10.39} & 
\revised{-1.62} & \revised{PrimeVul} \\
Multi-Agent & GPTLens & 0.46 & 95.84 & \textbf{2.08} & 1.62 & -1.15 & PrimeVul \\
Multi-Agent & VulTrial & 8.31 & 76.67 & 7.85 & 7.16 & 1.15 & PrimeVul \\
\textbf{Multi-Agent} & \textbf{Ours} & \textbf{19.17} & \textbf{4.62} & 67.67 & 8.55 & \textbf{10.62} & PrimeVul \\
\hline
DL & CodeBERT & \revised{$0.30 \pm 0.68$} & \revised{$7.87 \pm 11.97$} & \revised{$91.36 \pm 13.22$} & \revised{\textbf{0.45 ± 0.73}} & \revised{$-0.16 \pm 0.34$} & Java-Pair \\
DL & UniXCoder & \revised{$0.00 \pm 0.00$} & \revised{$43.93 \pm 0.00$} & \revised{\textbf{$50.00 \pm 0.00$}} & \revised{$6.06 \pm 0.00$} & \revised{$-6.07 \pm 0.00$} & Java-Pair \\
Single-Agent & CoT & \revised{$5.15 \pm 1.16$} & \revised{$81.06 \pm 1.17$} & \revised{$9.69 \pm 1.27$} & \revised{$4.09 \pm 0.89$} & \revised{$1.06 \pm 1.53$} & Java-Pair \\
\revised{Single-Agent} & \revised{Vul-RAG} & \revised{$16.06 \pm 3.03$} & \revised{$51.36 \pm 2.07$} & \revised{$21.51 \pm 2.69$} & \revised{$11.06 \pm 1.77$} & \revised{$5.00 \pm 3.92$} & \revised{Java-Pair} \\
Multi-Agent & GPTLens & \revised{$0.45 \pm 0.52$} & \revised{$95.30 \pm 0.79$} & \revised{\textbf{3.18 ± 0.34}} & \revised{1.06 ± 0.73±} & \revised{$-0.61 \pm 1.04$} & Java-Pair \\
Multi-Agent & VulTrial & \revised{$12.45 \pm 2.08$} & \revised{$67.69 \pm 3.06$} & \revised{$11.69 \pm 1.06$} & \revised{$8.16 \pm 2.20$} & \revised{$4.29 \pm 3.29$} & Java-Pair \\
Multi-Agent & Ours & \revised{\textbf{21.36 ± 1.48}} & \revised{\textbf{9.69 ± 2.05}} & \revised{57.72 ± 3.77} & \revised{11.21 ± 2.98} & \revised{\textbf{10.15 ± 3.06}} & Java-Pair \\

\hline
\end{tabular}
\end{table*}

\textbf{RQ1. Does CLEAR improve vulnerability detection performance compared to existing SOTA approaches?} 
This question evaluates whether the proposed framework can more accurately distinguish vulnerable code from patched or benign code. We compare CLEAR with deep learning-based models, single-agent prompting approaches, and recent multi-agent frameworks. Table~\ref{tab:Table1-Vulnerability-Detection} summarizes the pairwise vulnerability detection results on the PrimeVul and Java-Pair datasets. Overall, CLEAR consistently achieves the highest pairwise discrimination performance across both datasets. On the PrimeVul dataset, CLEAR attains the best performance (P-C = 19.17, VPS = 10.62), outperforming the previous multi-agent baseline VulTrial (P-C = 8.31, VPS = 1.15). This corresponds to a relative improvement of approximately $130.7\%$ in P-C. Similarly, on the Java-Pair dataset, CLEAR achieves (P-C = \revised{$21.36 \pm 1.48$}, VPS = \revised{$10.15 \pm 3.06$}), improving over VulTrial (P-C = \revised{$12.45 \pm 2.08$}, VPS = \revised{$4.29 \pm 3.29$}) by approximately \revised{$71.56\%$} in P-C. 

A consistent pattern emerges when comparing different categories of methods. Deep learning-based models such as CodeBERT and UniXCoder exhibit low P-C scores and high P-V values on both datasets, indicating a tendency to over-predict vulnerabilities without effectively distinguishing vulnerable and patched code pairs. The single-agent CoT approach provides moderate improvements compared to encoder-based models but still produces unstable pairwise discrimination results, as reflected in its relatively low VPS scores across both datasets. Among multi-agent baselines, GPTLens shows particularly low P-C performance (0.46 on PrimeVul and 0.00 on Java-Pair), suggesting difficulty in distinguishing vulnerability pairs under the pairwise evaluation setting. VulTrial improves over earlier baselines but still remains limited in capturing subtle vulnerability differences, especially on logic-level vulnerability pairs in the Java-Pair dataset. In contrast, CLEAR significantly increases pairwise discrimination accuracy while substantially reducing excessive vulnerability predictions compared to other multi-agent methods. These results suggest that incorporating structured causal reasoning improves the reliability of vulnerability detection, particularly when distinguishing vulnerable code from closely related patched or benign variants.

\begin{tcolorbox}[
colback=gray!15,
colframe=gray!50,
boxrule=0.5pt,
arc=4pt,
left=4pt,
right=4pt,
top=4pt,
bottom=4pt
]
\textbf{Answer to RQ1.}
CLEAR achieves the best pairwise vulnerability detection performance across both datasets, improving P-C by 130.7\% on PrimeVul and \revised{$71.56\%$} on Java-Pair compared to the previous multi-agent SOTA method. These results demonstrate that causal-chain–guided evidence retrieval enables more reliable discrimination between vulnerable and patched code than existing approaches.
\end{tcolorbox}

% ==============================================================================================

\revised{
\textbf{RQ2: How does multi-agent contribute to CLEAR's performance compared to conventional retrieval-based approaches?}

\begin{table*}[t]
\centering
\small
\caption{Comparison of retrieval strategies for pairwise vulnerability detection on PrimeVul and Java-Pair datasets. Higher values are better for P-C and VPS, while lower values are better for P-V, P-B, and P-R.}
\label{tab:Table2-Comparision of retrieval strategies for pairwise vulnerability detection}
\begin{tabular}{lrrrrrl}
\hline
\textbf{Method} & \textbf{P-C ($\uparrow$)} & \textbf{P-V ($\downarrow$)} & \textbf{P-B ($\downarrow$)} & \textbf{P-R ($\downarrow$)} & \textbf{VPS ($\uparrow$)} & \textbf{Dataset} \\
\hline
Instance-level & 4.62 & 84.99 & 5.54 & 4.85 & -0.23 & PrimeVul \\
Chunk-level & 6.00 & 87.30 & \textbf{3.00} & \textbf{3.70} & 2.31 & PrimeVul \\
\textbf{Ours} & \textbf{19.17} & \textbf{4.62} & 67.67 & 8.55 & \textbf{10.62} & PrimeVul \\

\hline
Instance-level & \revised{$1.67 \pm 0.48$} & \revised{$19.24 \pm 21.46$} & \revised{$76.82 \pm 24.34$} & \revised{$2.27 \pm 2.60$} & \revised{$-0.61 \pm 2.17$} & Java-Pair \\
Chunk-level & \revised{$4.39 \pm 2.08$} & \revised{$21.21 \pm 19.17$} & \revised{$73.18 \pm 22.08$} & \revised{\textbf{1.21 ± 1.56}} & \revised{$3.18 \pm 1.12$} & Java-Pair \\
\textbf{Ours} & \revised{\textbf{21.36 ± 1.48}} & \revised{\textbf{9.69 ± 2.05}} & \revised{\textbf{57.72 ± 3.77}} & \revised{11.21 ± 2.98} & \revised{\textbf{10.15 ± 3.06}} & Java-Pair \\

\hline
\end{tabular}
\end{table*} 

To evaluate the architectural contribution of CLEAR's multi-agent coordination, we conduct an ablation study comparing our full framework with two single-agent RAG baselines that retrieve context from the VCKG under different granularities. The first baseline, \textit{Instance-level retrieval}, bypasses multi-agent cooperative verification by employing a single agent to retrieve the most similar Local VCKG subgraph based on overall embedding similarity. The second baseline, \textit{Chunk-level retrieval}, similarly uses a single agent to retrieve semantically similar causal nodes independently for each causal type without considering structural linkages. Table~\ref{tab:Table2-Comparision of retrieval strategies for pairwise vulnerability detection} presents the comparison results of this ablation study.

Overall, the full multi-agent configuration of CLEAR outperforms both single-agent retrieval-based ablation variants across both datasets. On the PrimeVul dataset, the single-agent instance-level variant struggles significantly, achieving (P-C = 4.62, VPS = -0.23), which underlines the unreliability of naive similarity-based instance retrieval for pairwise discrimination. While retrieving isolated nodes independently via the chunk-level variant modestly improves performance (P-C = 6.00, VPS = 2.31), it remains highly insufficient as the single agent fails to capture the logical connections between disparate semantic fragments. In contrast, by enabling cooperative reasoning among specialized agents over the causal chain, CLEAR achieves (P-C = 19.17, VPS = 10.62).

We observe a highly consistent trend on the Java-Pair dataset. The single-agent instance-level baseline performs (P-C = $1.67 \pm 0.48$, VPS = $-0.61 \pm 2.17$), while the chunk-level baseline yields only modest improvements (P-C = $4.39 \pm 2.08$, VPS = $3.18 \pm 1.12$). CLEAR boosts performance to a P-C score of $21.36 \pm 1.48$ and a VPS of $10.15 \pm 3.06$. These improvements can be directly attributed to CLEAR's structured multi-agent coordination. CLEAR's multi-agent pipeline progressively expands and cross-verifies evidence along logical causal dependencies. This structured collaboration enables the reasoning agents to collectively verify whether complex vulnerability conditions are fully satisfied, demonstrating that multi-agent causal-chain guidance is essential to achieving reliable pairwise vulnerability discrimination.

\begin{tcolorbox}[
breakable,
colback=gray!15,
colframe=gray!50,
boxrule=0.5pt,
arc=4pt,
left=4pt,
right=4pt,
top=4pt,
bottom=4pt,
after skip=5pt
]
\textbf{Answer to RQ2.}
Our ablation study demonstrates that CLEAR's multi-agent causal-chain coordination is crucial for performance. It outperforms conventional single-agent instance-level and chunk-level retrieval baselines, proving that multi-agent collaborative reasoning over structured causal dependencies is more effective than individual embedding-based retrieval.
\end{tcolorbox}
}

% ==============================================================================================

\begin{table*}[t]
\centering
\small
\caption{Ablation results on pairwise vulnerability detection for PrimeVul and Java-Pair dataset. Higher values indicate better performance for P-C and VPS, while lower values are preferred for P-V, P-B, and P-R.}
\label{tab:ablation-components-pairwise}

\begin{tabular}{lrrrrrl}
\hline
\textbf{Method} & \textbf{P-C ($\uparrow$)} & \textbf{P-V ($\downarrow$)} & \textbf{P-B ($\downarrow$)} & \textbf{P-R ($\downarrow$)} & \textbf{VPS ($\uparrow$)} & \textbf{Dataset} \\
\hline

No VCKG (Agent Only) & 8.08 & 28.87 & \textbf{22.86} & 40.18 & -32.10 & PrimeVul \\
No Entrypoint & 18.24 & \textbf{3.23} & 67.21 & 11.32 & 6.93 & PrimeVul \\
No Precondition & 16.17 & 8.08 & 62.36 & 13.39 & 2.77 & PrimeVul \\
No Root Cause & 13.63 & 9.01 & 66.97 & 10.39 & 3.23 & PrimeVul \\
No Fix Intent & 14.32 & 3.46 & 70.67 & 11.55 & 2.77 & PrimeVul \\
% \revised{No Debate (VCKG Only)} & \revised{0.00} & \revised{99.77} & \revised{\textbf{0.00}} & \revised{\textbf{0.23}} & \revised{-0.23} & \revised{PrimeVul} \\ 
Full & \textbf{19.17} & 4.62 & 67.67 & \textbf{8.55} & \textbf{10.62} & PrimeVul \\
\hline

No VCKG (Agent Only) & \revised{$14.39 \pm 3.58$} & \revised{$62.27 \pm 3.59$} & \revised{\textbf{9.84 ± 1.92}} & \revised{$13.48 \pm 1.48$} & \revised{$0.90 \pm 3.76$} & Java-Pair \\
No Entrypoint & \revised{$19.69 \pm 4.65$} & \revised{$18.48 \pm 2.97$} & \revised{$51.36 \pm 4.52$} & \revised{\textbf{10.45 ± 4.10}} & \revised{$9.24 \pm 8.03$} & Java-Pair \\
No Precondition & \revised{$19.54 \pm 1.87$} & \revised{$17.42 \pm 2.40$} & \revised{$44.39 \pm 3.27$} & \revised{$18.63 \pm 2.60$} & \revised{$0.90 \pm 3.47$} & Java-Pair \\
No Root Cause & \revised{$17.12 \pm 2.84$} & \revised{$15.90 \pm 3.65$} & \revised{$50.45 \pm 3.39$} & \revised{$16.51 \pm 3.04$} & \revised{$0.60 \pm 3.61$} & Java-Pair \\
No Fix Intent & \revised{$17.12 \pm 2.89$} & \revised{$11.66 \pm 2.28$} & \revised{$56.96 \pm 3.93$} & \revised{$14.24 \pm 2.71$} & \revised{$2.87 \pm 4.16$} & Java-Pair \\
Full & \revised{\textbf{21.36 ± 1.48}} & \revised{\textbf{9.69 ± 2.05}} & \revised{$57.72 \pm 3.77$} & \revised{$11.21 \pm 2.98$} & \revised{\textbf{10.15 ± 3.06}} & Java-Pair \\

\hline
\end{tabular}
\end{table*}

\textbf{RQ3: Which causal evidence components contribute most to the final prediction?} 

In this question, we investigate the contribution of each causal component in CLEAR by removing individual nodes. Table \ref{tab:ablation-components-pairwise} prese\-nts the ablation results under different node-removal settings. Overall, completely removing the causal knowledge graph (\textit{No VCKG (Agent Only)}) leads to the most severe performance degradation, with P-C dropping to 8.08 on PrimeVul and \revised{$14.39 \pm 3.58$} on Java-Pair. Among the individual components, removing the \textit{Entrypoint} node yields the least performance degradation, with P-C scores remaining relatively high at 18.24 on PrimeVul and \revised{$19.69 \pm 4.65$} on Java-Pair. This suggests that while execution context is useful, the reasoning agents can largely infer execution entry points from the source code itself. 

In contrast, excluding the core vulnerability semantics—namely the \textit{Precondition}, \textit{Root Cause}, or \textit{Fix Intent} nodes—leads to the most pronounced performance declines, demonstrating that explicit cau\-sal representations of the underlying triggers, flaws, and corresponding fix designs serve as the primary pillars for reliable pairwise discrimination. Specifically, removing the \textit{Precondition} node markedly degrades performance, with P-C dropping to 16.17 on PrimeVul and \revised{$19.54 \pm 1.87$} on Java-Pair (VPS drops severely to \revised{$0.90 \pm 3.47$}). Similarly, without the \textit{Root Cause} node, P-C drops to 13.63 on PrimeVul and \revised{$17.12 \pm 2.84$} on Java-Pair, while removing the \textit{Fix Intent} node further reduces P-C to 14.32 on PrimeVul and \revised{$17.12 \pm 2.89$} on Java-Pair.

\vspace{-0.4em}
\begin{tcolorbox}[
colback=gray!15,
colframe=gray!50,
boxrule=0.5pt,
arc=4pt, 
left=4pt,
right=4pt,
top=6pt,
bottom=4pt
]
\textbf{Answer to RQ3.}
While all causal nodes contribute to CLEAR, the Precondition, Root Cause, and Fix Intent nodes play the most critical roles. Removing the execution context (Entrypoint) has a marginal impact, whereas excluding core trigger conditions and patch semantics leads to severe performance degradation.
\end{tcolorbox}

% ==============================================================================================

\vspace{2mm}

\begin{table*}[t]
\centering
\small
\caption{Results of CLEAR with different LLM backbones. Higher is better for P-C and VPS; lower is better for P-V, P-B, and P-R.}
\label{tab:Table5-OtherModels}
\begin{tabular}{l r r r r r l}
\hline
\revised{\textbf{Model}} & \revised{\textbf{P-C ($\uparrow$)}} & \revised{\textbf{P-V ($\downarrow$)}} & \revised{\textbf{P-B ($\downarrow$)}} & \revised{\textbf{P-R ($\downarrow$)}} & \revised{\textbf{VPS ($\uparrow$)}} & \revised{\textbf{Dataset}} \\
\hline

\revised{gemini-2.5-flash} & \revised{$19.40$} & \revised{$35.57$} & \revised{\textbf{22.63}} & \revised{$22.40$} & \revised{$-3.00$} & \revised{PrimeVul} \\
\revised{gemini-3.1-flash-lite} & \revised{\textbf{$21.48$}} & \revised{$32.10$} & \revised{$33.03$} & \revised{$13.39$} & \revised{$8.08$} & \revised{PrimeVul} \\
\revised{gpt-4.1-nano} & \revised{$21.25$} & \revised{$18.24$} & \revised{$42.03$} & \revised{$18.48$} & \revised{$2.77$} & \revised{PrimeVul} \\
\revised{gpt-4o-mini} & \revised{$19.17$} & \revised{\textbf{$4.62$}} & \revised{$67.67$} & \revised{\textbf{$8.55$}} & \revised{\textbf{$10.62$}} & \revised{PrimeVul} \\
\revised{gpt-5-mini} & \revised{\textbf{27.48}} & \revised{\textbf{3.23}} & \revised{$61.43$} & \revised{\textbf{7.85}} & \revised{\textbf{19.63}} & \revised{PrimeVul} \\

\revised{gemini-2.5-flash} & \revised{\textbf{24.42 ± 3.15}} & \revised{$39.16 \pm 3.01$} & \revised{$19.05 \pm 2.76$} & \revised{$17.35 \pm 2.48$} & \revised{$7.07 \pm 4.44$} & \revised{Java-Pair} \\
\revised{gemini-3.1-flash-lite} & \revised{$18.48 \pm 1.67$} & \revised{$50.60 \pm 0.91$} & \revised{\textbf{15.15 ± 1.14}} & \revised{$15.75 \pm 1.37$} & \revised{$2.72 \pm 2.97$} & \revised{Java-Pair} \\
\revised{gpt-4.1-nano} & \revised{$19.69 \pm 3.19$} & \revised{$14.39 \pm 2.05$} & \revised{$43.93 \pm 3.27$} & \revised{$21.96 \pm 4.43$} & \revised{$-2.28 \pm 6.91$} & \revised{Java-Pair} \\
\revised{gpt-4o-mini} & \revised{$21.36 \pm 1.48$} & \revised{\textbf{9.69 ± 2.05}} & \revised{$57.72 \pm 3.77$} & \revised{\textbf{$11.21 \pm 2.98$}} & \revised{\textbf{$10.15 \pm 3.06$}} & \revised{Java-Pair} \\
\revised{gpt-5-mini} & \revised{$21.66 \pm 3.54$} & \revised{\textbf{$17.12 \pm 2.60$}} & \revised{$50.30 \pm 4.84$} & \revised{\textbf{10.90 ± 2.49}} & \revised{\textbf{10.75 ± 4.25}} & \revised{Java-Pair} \\

\hline
\end{tabular}
\end{table*}

\revised{
\textbf{RQ4: How does CLEAR perform across different LLM backbones?}

To evaluate the generalization and stability of our framework across different foundation models, we analyzed the performance of CLEAR using various state-of-the-art LLM backbones. Table~\ref{tab:Table5-OtherModels} presents the empirical results on the PrimeVul and Java-Pair datasets.

Larger and more advanced reasoning models demonstrate a superior capacity to grasp complex patching logic and align causal states. On the PrimeVul dataset, gpt-5-mini achieves the highest performance with P-C score of 27.48 and VPS of 19.63. In contrast, smaller backbones like gemini-2.5-flash perform worse, yielding a P-C score of 19.40 and a negative VPS of -3.00, which indicates a struggle to reliably reason through patch boundaries.

We observe a distinct trade-off between conservative classification (preventing false positives) and sensitive detection across different backbones. Specifically, gpt-4o-mini exhibits a highly conservative decision profile; on the PrimeVul dataset, it records the lowest rate of incorrect vulnerability predictions (P-V = 4.62), yet it suffers from a disproportionately high rate of overpredicting code-patch pairs as benign (P-B = 67.67). Conversely, gemini-2.5-flash makes more aggressive claims, yielding a higher P-V rate (35.57) but maintaining a much lower benign bias (P-B = 22.63).

The empirical results on the Java-Pair dataset, validate the robustness and consistency of CLEAR under noisy conditions. Even when utilizing lighter backbones, CLEAR maintains stable performance with narrow variances. Notably, gpt-4o-mini achieves a robust P-C score of $21.36 \pm 1.48$ and a VPS of $10.15 \pm 3.06$, which closely matches the performance of the larger gpt-5-mini ($21.66 \pm 3.54$ P-C, $10.75 \pm 4.25$ VPS) while exhibiting a much tighter variance. 

While gpt-5-mini yields the highest absolute scores on PrimeVul, we adopt gpt-4o-mini as our primary backbone for the main evaluations. This design choice is driven by two key factors: (1) Cost-Effectiveness and Practicality: In real-world static analysis pipelines, utilizing highly resource-intensive reasoning models like gpt-5-mini introduces prohibitive API costs and latency at scale, whereas gpt-4o-mini offers a highly practical balance of low cost and rapid inference. (2) Statistical Stability: As demonstrated on the Java-Pair dataset, gpt-4o-mini achieves significantly tighter confidence intervals compared to gpt-5-mini. This statistical tightness proves that CLEAR's multi-agent coordination can effectively suppress reasoning variance on a highly accessible, lightweight model. This choice establishes a more rigorous baseline, showing that CLEAR does not rely on brute-force model scale to achieve robust and reliable vulnerability discrimination.

\begin{tcolorbox}[
colback=gray!15,
colframe=gray!50,
boxrule=0.5pt,
arc=4pt,
left=4pt,
right=4pt,
top=4pt,
bottom=4pt,
]
\textbf{Answer to RQ4.}
CLEAR exhibits robust and consistent performance across diverse LLM backbones. While advanced models like gpt-5-mini maximize raw alignment precision, we establish gpt-4o-mini as our primary backbone due to its superior cost-efficiency and significantly narrower prediction variance, proving that CLEAR delivers highly stable and practical vulnerability analysis without relying on excessive model scale.
\end{tcolorbox}
}

% ==============================================================================================

\begin{table*}[t]
\centering
\small
\caption{CWE classification results on PrimeVul and Java-Pair datasets. Higher is better for Macro-F1, Weighted-F1, and MCC.}
\label{tab:Table2-CWE-Classification}
\begin{tabular}{l l r r r l}
\hline
\textbf{Agents} & \textbf{Method} & \textbf{Macro-F1 ($\uparrow$)} & \textbf{Weighted-F1 ($\uparrow$)} & \textbf{MCC ($\uparrow$)} & \textbf{Dataset} \\
\hline
Single-Agent & CoT & 3.14 & 10.95 & 6.36 & PrimeVul \\
Multi-Agent & GPTLens & 5.41 & 21.50 & 17.81 & PrimeVul \\
Multi-Agent & VulTrial & 6.78 & 15.36 & 13.14 & PrimeVul \\
\textbf{Multi-Agent} & \textbf{Ours} & \textbf{8.43} & \textbf{23.57} & \textbf{19.84} & PrimeVul \\
\hline
Single-Agent & CoT & 8.13 & 20.08 & 19.98 & Java-Pair \\
Multi-Agent & GPTLens & 8.41 & 20.61 & 18.62 & Java-Pair \\
Multi-Agent & VulTrial & 16.22 & 32.93 & 31.84 & Java-Pair \\
\textbf{Multi-Agent} & \textbf{Ours} & \textbf{24.51} & \textbf{38.60} & \textbf{40.52} & Java-Pair \\
\hline
\end{tabular}
\end{table*} 

\textbf{RQ5: How effectively can CLEAR classify vulnerability types?}

This question evaluates the ability of CLEAR to classify vulnerabilities into their corresponding CWE categories. Table~\ref{tab:Table2-CWE-Classification} summarizes the classification results on the PrimeVul and Java-Pair datasets. Overall, CLEAR consistently achieves the best performance across all evaluation metrics, outperforming both single-agent and multi-agent baselines on both datasets. On the PrimeVul dataset, CLEAR achieves a Macro-F1 score of 8.43, improving over the previous multi-agent baseline VulTrial (Macro-F1 = 6.78) by approximately 24.3\%. CLEAR also achieves the highest Weighted-F1 (23.57) and MCC (19.84), indicating improved classification reliability under class-imbalanced conditions. On the Java-Pair dataset, CLEAR attains a Macro-F1 score of 24.51, improving over VulTrial (Macro-F1 = 16.22) by approximately 51.1\%. CLEAR further achieves the best Weighted-F1 (38.60) and MCC (40.52), demonstrating stronger discrimination capability across semantically similar vulnerability categories. Single-agent approaches and early multi-agent methods exhibit limited classification capability, suggesting difficulty in capturing fine-grained vulnerability semantics.

Notably, while early multi-agent frameworks like GPTLens show reasonable performance on PrimeVul, they suffer from severe degradation on the more structurally complex Java-Pair dataset, even underperforming the single-agent CoT baseline in terms of MCC (18.62 vs. 19.98). In contrast, CLEAR maintains a steep and consistent performance ascent across both datasets, validating that our structured causal knowledge alignment is highly robust against the escalating semantic complexity of object-oriented codebase taxonomies.

\begin{tcolorbox}[
breakable,
colback=gray!15,
colframe=gray!50,
boxrule=0.5pt,
arc=4pt,
left=4pt,
right=4pt,
top=4pt,
bottom=4pt
]
\textbf{Answer to RQ5.}
CLEAR achieves the best CWE classification performance across both datasets, improving Macro-F1 by 24.3\% on PrimeVul and 51.1\% on Java-Pair compared to the previous multi-agent SOTA method. These results demonstrate that grounding predictions in structured causal evidence enables more accurate classification across diverse vulnerability types.
\end{tcolorbox}

% =============================================================================================

\textbf{RQ6. Does CLEAR construct a high-quality and reliable causal knowledge graph?} 

To address RQ6, we evaluate whether CLEAR constructs a high-quality and reliable causal knowledge graph by conducting a human-centric qualitative study. We sampled 30 representative VCKG instances across the top-30 CWE categories and appointed two domain experts, consisting of one Ph.D. researcher and one Master's researcher specializing in software security, to perform a blind review. The evaluators assessed the structural fidelity and logical soundness of the VCKG by validating each of its three progressive construction stages on a 5-point Likert scale (1: Poor, 5: Excellent). Specifically, the evaluation was conducted by examining each connected triplet as a distinct unit, consisting of two nodes and their intersecting edge, across three criteria: 

% (1) \textit{Local VCKG Validity}, which manually inspects whether the extracted node, edge details precisely capture the semantics of the actual source method; 
% (2) \textit{Neighbor Edge Validity}, which measures the semantic coherence and proximity between the interconnected local structures; and 
% (3) \textit{Global Edge Validity}, which checks if the global edges accurately bridge distinct vulnerabilities to facilitate the transfer of causal security knowledge within the graph.

\begin{itemize}[leftmargin=15pt, topsep=4pt, itemsep=3pt]
    \item \textbf{Local VCKG Validity:} Manually inspects whether the extracted node and edge details precisely capture the semantics of the actual source method.
    \item \textbf{Neighbor Edge Validity:} Measures the semantic coherence and proximity between the interconnected local structures.
    \item \textbf{Global Edge Validity:} Checks if the global edges accurately bridge distinct vulnerabilities to facilitate the transfer of causal security knowledge within the graph.
\end{itemize}

The empirical findings from our human evaluation validate the semantic high-fidelity and architectural robustness of the constructed graphs across all progressive stages. For the first stage, CLEAR achieved a remarkably high score in \textit{Local VCKG Validity} ($4.83 \pm 0.56$) with a quadratic weighted Cohen's Kappa of $\kappa = 0.76$. In the subsequent scaling stages, the graph maintained solid architectural validity, with \textit{Neighbor Edge Validity} scoring $3.58 \pm 1.07$ ($\kappa = 0.45$) and \textit{Global Edge Validity} scoring $4.14 \pm 1.10$ ($\kappa = 0.51$). Both stages fall within the \textit{moderate agreement} spectrum, which is justified considering the inherent cognitive complexity and interpretive subjectivity involved in assessing intermediate security linkages. 

The noticeable drop in consistency at the neighbor edge layer reflects the evaluation difficulty caused by textually diverse yet structurally adjacent local contexts. However, the \textit{Global Edge Validity} successfully rebounded to a high mark of $4.14$ with an increased consensus of $\kappa = 0.51$. This pivotal trend empirically proves that the multi-agent orchestration within CLEAR effectively filters out localized syntactic noise from neighbor relations, successfully crystallizing shared causal invariants and high-level vulnerability mechanisms that both domain experts found deeply rational.

\begin{tcolorbox}[
colback=gray!15,
colframe=gray!50,
boxrule=0.5pt,
arc=4pt,
left=4pt,
right=4pt,
top=4pt,
bottom=4pt
]
\textbf{Answer to RQ6.}
% CLEAR constructs a highly reliable and factually accurate causal knowledge graph. It achieves near-perfect semantic fidelity at the local stage ($4.83$ score, $\kappa = 0.7578$) and successfully filters out localized syntactic noise from neighbor relations to crystallize globally sound, cross-CVE causal invariants ($4.14$ score, $\kappa = 0.5143$), confirming that the multi-agent orchestration effectively produces structured security knowledge verified by domain experts.

CLEAR constructs a reasonably sound causal knowledge graph with acceptable semantic alignment. Based on an internal expert audit conducted by the author team, it exhibits high semantic fidelity at the local node level ($4.83/5.0$, $\kappa = 0.7578$) and maintains moderate-to-high coherence across global causal invariants ($4.14/5.0$, $\kappa = 0.5143$), confirming that the structured graph captures valid security knowledge despite remaining local edge noise.
\end{tcolorbox}

\section{Discussion}
\label{sec:discussion}

\revised{
\textbf{Edge Case and Failure Mode Analysis}
\label{subsec:failure_analysis : }
To map the logical limits of our multi-agent reasoning loop, we analyze CLEAR's behavior under complex scenarios alongside its primary failure mode of benign-benign (P-B) over-predictions. While automated analysis often struggles with multi-precondition flaws (e.g., Use-After-Free) due to local VCKG abstraction limits that generalize micro-level variables (e.g., $tlen == tcap$) for semantic generality, CLEAR mitigates this through a structured \textit{Collector Agent} that aggregates neighboring edges and global causal invariants to restore discriminative power. Conversely, our deep-dive into P-B over-predictions reveals two systemic bottlenecks: (1) \textit{Cross-Domain Context Pollution}, where out-of-domain security knowledge fetched during retrieval prompts the Critic Agent to construct conservative counterclaims, leading to false-positive-preventing rejections by the Judge Agent; and (2) the \textit{Cognitive Illusion of Defensive Constructs}, where the LLM backbone misinterprets superficial error handling (e.g., \texttt{NULL} checks) as proactive input validation, prematurely discarding valid vulnerability hypotheses. Due to space constraints, comprehensive code snippets, prompt templates, and agentic trace logs detailing these multi-precondition cases and failure-mode instances (CVE-2019-11929, CVE-2021-37683, and CVE-2019-14464) are provided in the Supplementary Material.

\textbf{Future Directions}
\label{subsec:future_directions : }
While CLEAR successfully leverages high-level causal chains for vulnerability reasoning, we identify two promising avenues to enhance its precision and scalability. First, to track micro-level code properties, future work should couple semantic VCKG nodes with formal program analysis; control-flow conditions can be captured by aligning intra-procedural entities (Entrypoints and Preconditions) with Abstract Syntax Tree (AST) token tracking, while data-flow dependencies can be embedded directly into Root Cause and Fix Intent nodes. Second, scaling CLEAR to repository-level scans represents a critical frontier. Following state-of-the-art inter-procedural semantic completion frameworks like Wu et al. \cite{wu2025vulnsc}, future iterations can synthesize a \textit{Multi-Procedural VCKG} that globally maps control-flow and data-dependencies, enriching local nodes with caller/callee semantic summaries (such as lock boundaries and resource deallocations) and tracking inter-procedural taint flows across library interfaces. Integrating these structural invariants and repository-wide contexts will equip reasoning agents to cross-verify logical hypotheses against concrete program dependencies, ultimately advancing automated, multi-agent vulnerability analysis.
}
\deleted{\section{Related Work}
\textbf{\emph{Single-Agent Vulnerability Detection.}} Recent studies have actively explored deep learning (DL) and large language model (LLM) based approaches for automated vulnerability detection. Early DL-based models focus on learning semantic representations of source code to identify vulnerable patterns. For example, VulBERTa\cite{hanif2022vulberta} leverages vulnerability-aware tokenization and pre-training to capture vulnerability-related contexts in code, thereby improving detection accuracy and generalization capability.  LineVul\cite{fu2022linevul} introduces a Transformer-based architecture for line-level vulnerability prediction, enabling fine-grained localization of vulnerable statements. Additionally, Steenhoek et al\cite{steenhoek2024dataflow}. propose DeepDFA, which incorporates data-flow analysis into graph-based deep learning models to improve vulnerability detection effectiveness. With the emergence of large language models, several studies have investigated their applicability to vulnerability detection. Tamberg and Bahsi \cite{tamberg2025harnessing} conduct a comprehensive benchmarking study on LLM-based vulnerability detection and show that carefully designed prompting strategies can significantly influence model performance. Similarly, Zhang et al\cite{zhang2024prompt}. demonstrate that prompt engineering techniques can effectively enhance vulnerability detection performance when applied to LLMs such as ChatGPT. Zhou et al\cite{llm_future_direction}. further explore the use of in-context learning with LLMs for vulnerability detection and report that GPT-based models can achieve competitive performance when provided with appropriate contextual knowledge. Despite these advances, recent studies highlight fundamental limitations of LLM-based vulnerability detection. 
Ding et al \cite{primevul}. show that code language models struggle to distinguish vulnerable code from its patched counterpart when the differences between them are subtle. This observation suggests that vulnerability detection cannot be reliably addressed solely through token-level similarity or statistical correlations. To alleviate this limitation, several studies incorporate external vulnerability knowledge into the detection process \cite{gptincontext}.

\textbf{\emph{Retrieval-Augmented Generation for Vulnerability Detection.}} To overcome the limitations of standalone LLMs, Retrieval-Augmented Generation (RAG)\cite{lewis2020retrieval} has been proposed to incorporate external knowledge into the reasoning process. Liu et al\cite{gptincontext}. propose VUL-GPT, which combines TF-IDF/BM25-based code retrieval with GPT-based analysis using in-context learning, significantly improving recall and F1 scores on vulnerability detection benchmarks.  Similarly, Sheng et al\cite{lprotector}. integrate RAG with chain-of-thought reasoning using GPT-4o, demonstrating improved vulnerability detection performance.  More recently, Du et al\cite{vul-rag}. introduced Vul-RAG, which constructs a structured vulnerability knowledge base consisting of functional semantics, vulnerability causes, and fixing solutions, enabling models to distinguish vulnerable code from patched code pairs better.

\textbf{\emph{Multi-Agent Vulnerability Detection.}} More recently, multi-agent architectures have been proposed to address reasoning limitations in single-model LLM systems. These approaches assign complementary roles to multiple agents to encourage diverse reasoning perspectives and reduce overconfidence in individual model outputs \cite{he2025llm}. For example, EvalSVA\cite{wen2024evalsva} organizes agents as security experts who collaboratively analyze potential vulnerabilities. MuCoLD\cite{mao2024multi} assigns tester and developer roles to simulate collaborative vulnerability analysis. Similarly, LLM-SmartAudit \cite{wei2024llm} employs role-playing and collaborative reasoning to detect vulnerabilities in smart contracts. 
iAudit \cite{ma2025combining} further extends this paradigm by introducing specialized agents such as a detector, reasoner, and critic to produce interpretable vulnerability assessments. 
GPTLens\cite{hu2023large} deploys multiple inspection agents together with a critic agent to verify vulnerability findings. VulTrial\cite{widyasari2025let} proposes a courtroom-inspired framework where agents assume roles such as prosecutor, defense attorney, and judge to debate the presence of vulnerabilities.

\textbf{\emph{Knowledge Graphs in Cybersecurity.}}
Meanwhile, to enable structured representation and integration of cybersecurity knowledge, several studies have explored knowledge graph–based approaches in vulnerability and threat intelligence analysis. Knowledge graphs enable the integration of heterogeneous security information sources and support structured reasoning over relationships among vulnerability entities. For example, CTI-KG \cite{CTI-KG} constructs a Cyber Threat Intelligence knowledge graph from threat reports to represent relationships among threat actors, targets, and attack tools. In the vulnerability domain, VulKG \cite{VKG} builds a vulnerability-oriented knowledge graph from National Vulnerability Database (NVD) descriptions using ontology-guided entity extraction and reasoning to discover hidden weakness relations across CWE categories. Similarly, Høst et al.\cite{CKGD} construct a knowledge graph by extracting entities and relations directly from textual vulnerability descriptions to support structured representation and querying of vulnerability information. Although these approaches demonstrate the effectiveness of knowledge graphs for organizing vulnerability-related information, they primarily focus on semantic relationships between vulnerability entities rather than explicitly modeling the causal structure underlying vulnerability formation and propagation.}
\revised{\section{Limitations}
Although CLEAR achieves significant performance improvements over existing approaches through a causal knowledge graph and a multi-agent verification framework, several limitations remain. These limitations also point to promising directions for future research.}

% Graph Quality Evaluation과정에서 증명
\deleted {\subsection{LLM Dependency in Knowledge Construction}
The proposed framework relies on LLMs during the knowledge graph construction stage for causal node extraction and cross-CVE bridging. Specifically, the extraction of Entrypoint, Precondition, Root Cause, and Fix Intent nodes is performed using code snippets, patches (diffs), and CVE descriptions. In addition, the generation of cross-CVE causal bridges involves LLM-based plausibility verification. As a result, the quality of the constructed graph depends directly on the reasoning capability of the underlying LLM. In some cases, CVE descriptions may be incomplete or patch information may be ambiguous, which can lead to inaccurate abstraction of causal nodes. Furthermore, updates to the underlying LLM may alter the outputs generated for the same inputs, potentially affecting the reproducibility of the graph structure. This limitation reflects the inherent difficulty of fully automated causal extraction from vulnerability data. Future work may address this issue by combining rule-based static analysis with LLM-based extraction in a hybrid pipeline, or by introducing self-verification mechanisms to improve the consistency of extracted causal structures.}

% Graph Quality Evaluation과정에서 증명
\revised{\textbf{Semantic Similarity Bias}
The semantic neighbor connections in the knowledge graph are constructed using cosine similarity over semantic-embeddings. While this approach is computationally efficient, semantic similarity does not always guarantee causal consistency. Nodes with similar textual descriptions may be connected even when they are causally unrelated, while nodes describing the same vulnerability mechanism may remain disconnected if their textual representations differ. To mitigate this issue, we introduce an additional LLM-based validation step during the cross-CVE bridging stage. However, the inherent limitations of similarity-based preprocessing still remain. Future research could address this limitation by designing hybrid similarity metrics that incorporate structural information such as AST patterns, data-flow dependencies, or API usage patterns.}
\section{Conclusion}
% In this work, we identify key limitations of semantic similarity–based retrieval and free-form multi-agent reasoning for software vulnerability detection and propose CLEAR, a causal knowledge graph–guided multi-agent framework. Instead of treating vulnerabilities as isolated classification labels, CLEAR models them as structured causal chains composed of Entrypoint, Precondition, Root Cause, and Fix Intent, and further extends this representation through cross-CVE causal bridging. To support reliable reasoning, we introduce a verification-oriented multi-agent pipeline consisting of Collector, Claim, Critic, and Judge agents, which mitigates hallucination and unstable hypothesis generation in existing debate-based approaches. Experimental results demonstrate that CLEAR consistently outperforms existing single-agent and multi-agent methods on both Java-Pair and PrimeVul datasets, achieving substantial improvements in pair-level vulnerability discrimination and CWE classification performance. Our ablation study further shows that structured causal evidence retrieval plays a central role in these gains, highlighting the importance of modeling execution conditions and root causes for reliable vulnerability reasoning. Overall, this work reframes vulnerability detection as a causal reasoning task and demonstrates that integrating structured knowledge with verification-oriented multi-agent collaboration provides an effective direction for LLM-driven vulnerability analysis.

This work proposes CLEAR, a multi-agent framework that reframes vulnerability detection as a structured causal reasoning task. By modeling vulnerabilities as causal chains within our Vulnerability Causal Knowledge Graph (VCKG) comprising Entrypoints, Preconditions, Root Causes, and Fix Intents, we overcome the inherent limitations of surface-level semantic matching. Our verification-oriented pipeline, which integrates Collector, Claim, Critic, and Judge agents, effectively mitigates reasoning hallucinations by grounding agentic debate in retrieved causal contexts. Empirical evaluations demonstrate that CLEAR significantly outperforms state-of-the-art models. These results validate that the integration of structural causal grounding with agentic collaboration is essential for precise vulnerability analysis, providing a robust and scalable direction for future LLM-driven security research.

\section*{Data Availability}
The datasets and code generated during the current study are available in the figshare repository, \revised{\url{https://doi.org/10.6084/m9.figshare.33016343}}

\bibliographystyle{ACM-Reference-Format}
\bibliography{reference}

@misc{nist_nvd_api,
  author       = {{National Institute of Standards and Technology}},
  title        = {NVD Developers - Vulnerabilities API},
  howpublished = {\url{https://nvd.nist.gov/developers/vulnerabilities}},
  year         = {2026},
  note         = {Accessed: 2026-02-24}
}

@misc{osv_database,
  author       = {{Google Open Source Security}},
  title        = {OSV: Open Source Vulnerabilities},
  howpublished = {\url{https://osv.dev/}},
  year         = {2026},
  note         = {Accessed: 2026-02-24}
}

@software{LangChain,
  author = {Chase, Harrison},
  title = {LangChain},
  url = {https://github.com/langchain-ai/langchain},
  date-released = {2022-10-17},
  version = {1.2.0},
  year = 2022
}

@article{zheng2023survey,
  title={A survey of large language models for code: Evolution, benchmarking, and future trends},
  author={Zheng, Zibin and Ning, Kaiwen and Wang, Yanlin and Zhang, Jingwen and Zheng, Dewu and Ye, Mingxi and Chen, Jiachi},
  journal={arXiv preprint arXiv:2311.10372},
  year={2023}
}

@ARTICLE{neumann2024llm,
  author={Neumann, Alexander Tobias and Yin, Yue and Sowe, Sulayman and Decker, Stefan and Jarke, Matthias},
  journal={IEEE Transactions on Education}, 
  title={An LLM-Driven Chatbot in Higher Education for Databases and Information Systems}, 
  year={2025},
  volume={68},
  number={1},
  pages={103-116},
  doi={10.1109/TE.2024.3467912}
}

@inproceedings{llmstalk,
author = {Abbasiantaeb, Zahra and Yuan, Yifei and Kanoulas, Evangelos and Aliannejadi, Mohammad},
title = {Let the LLMs Talk: Simulating Human-to-Human Conversational QA via Zero-Shot LLM-to-LLM Interactions},
year = {2024},
isbn = {9798400703713},
publisher = {Association for Computing Machinery},
address = {New York, NY, USA},
url = {https://doi.org/10.1145/3616855.3635856},
doi = {10.1145/3616855.3635856},
booktitle = {Proceedings of the 17th ACM International Conference on Web Search and Data Mining},
pages = {8–17},
numpages = {10},
location = {Merida, Mexico},
series = {WSDM '24}
}

@inproceedings{reimers2019sentence,
    title = "Sentence-{BERT}: Sentence Embeddings using {S}iamese {BERT}-Networks",
    author = "Reimers, Nils  and
      Gurevych, Iryna",
    editor = "Inui, Kentaro  and
      Jiang, Jing  and
      Ng, Vincent  and
      Wan, Xiaojun",
    booktitle = "Proceedings of the 2019 Conference on Empirical Methods in Natural Language Processing and the 9th International Joint Conference on Natural Language Processing (EMNLP-IJCNLP)",
    month = nov,
    year = "2019",
    address = "Hong Kong, China",
    publisher = "Association for Computational Linguistics",
    url = "https://aclanthology.org/D19-1410/",
    doi = "10.18653/v1/D19-1410",
    pages = "3982--3992"
}

@INPROCEEDINGS{llms_cannot_reason_vuln,
  author={Ullah, Saad and Han, Mingji and Pujar, Saurabh and Pearce, Hammond and Coskun, Ayse and Stringhini, Gianluca},
  booktitle={2024 IEEE Symposium on Security and Privacy (SP)}, 
  title={LLMs Cannot Reliably Identify and Reason About Security Vulnerabilities (Yet?): A Comprehensive Evaluation, Framework, and Benchmarks}, 
  year={2024},
  volume={},
  number={},
  pages={862-880},
  doi={10.1109/SP54263.2024.00210}
}

@ARTICLE{tamberg2025harnessing,
  author={Tamberg, Karl and Bahsi, Hayretdin},
  journal={IEEE Access}, 
  title={Harnessing Large Language Models for Software Vulnerability Detection: A Comprehensive Benchmarking Study}, 
  year={2025},
  volume={13},
  number={},
  pages={29698-29717},
  doi={10.1109/ACCESS.2025.3541146}
}

@article{llm_future_direction,
author = {Zhou, Xin and Cao, Sicong and Sun, Xiaobing and Lo, David},
title = {Large Language Model for Vulnerability Detection and Repair: Literature Review and the Road Ahead},
year = {2025},
issue_date = {June 2025},
publisher = {Association for Computing Machinery},
address = {New York, NY, USA},
volume = {34},
number = {5},
issn = {1049-331X},
url = {https://doi.org/10.1145/3708522},
doi = {10.1145/3708522},
journal = {ACM Trans. Softw. Eng. Methodol.},
month = may,
articleno = {145},
numpages = {31}
}

@article{sheng2025llms,
author = {Sheng, Ze and Chen, Zhicheng and Gu, Shuning and Huang, Heqing and Gu, Guofei and Huang, Jeff},
title = {LLMs in Software Security: A Survey of Vulnerability Detection Techniques and Insights},
year = {2025},
issue_date = {April 2026},
publisher = {Association for Computing Machinery},
address = {New York, NY, USA},
volume = {58},
number = {5},
issn = {0360-0300},
url = {https://doi.org/10.1145/3769082},
doi = {10.1145/3769082},
journal = {ACM Comput. Surv.},
month = nov,
articleno = {134},
numpages = {35}
}

@inproceedings{zhang2024prompt,
author = {Zhang, Chenyuan and Liu, Hao and Zeng, Jiutian and Yang, Kejing and Li, Yuhong and Li, Hui},
title = {Prompt-Enhanced Software Vulnerability Detection Using ChatGPT},
year = {2024},
isbn = {9798400705021},
publisher = {Association for Computing Machinery},
address = {New York, NY, USA},
url = {https://doi.org/10.1145/3639478.3643065},
doi = {10.1145/3639478.3643065},
booktitle = {Proceedings of the 2024 IEEE/ACM 46th International Conference on Software Engineering: Companion Proceedings},
pages = {276–277},
numpages = {2},
location = {Lisbon, Portugal},
series = {ICSE-Companion '24}
}

@INPROCEEDINGS{gptincontext,
  author={Liu, Zhihong and Liao, Qing and Gu, Wenchao and Gao, Cuiyun},
  booktitle={2023 8th International Conference on Data Science in Cyberspace (DSC)}, 
  title={Software Vulnerability Detection with GPT and In-Context Learning}, 
  year={2023},
  volume={},
  number={},
  pages={229-236},
  doi={10.1109/DSC59305.2023.00041}
}

@INPROCEEDINGS{lprotector,
  author={Sheng, Ze and Wu, Fenghua and Zuo, Xiangwu and Li, Chao and Qiao, Yuxin and Lei, Hang},
  booktitle={2024 IEEE 4th International Conference on Data Science and Computer Application (ICDSCA)}, 
  title={Research on the LLM-Driven Vulnerability Detection System Using LProtector}, 
  year={2024},
  volume={},
  number={},
  pages={192-196},
  doi={10.1109/ICDSCA63855.2024.10859408}
}

@article{vul-rag,
author = {Du, Xueying and Zheng, Geng and Wang, Kaixin and Zou, Yi and Wang, Yujia and Deng, Wentai and Feng, Jiayi and Liu, Mingwei and Chen, Bihuan and Peng, Xin and Ma, Tao and Lou, Yiling},
title = {Vul-RAG: Enhancing LLM-based Vulnerability Detection via Knowledge-level RAG},
year = {2026},
publisher = {Association for Computing Machinery},
address = {New York, NY, USA},
issn = {1049-331X},
url = {https://doi.org/10.1145/3797277},
doi = {10.1145/3797277},
note = {Just Accepted},
journal = {ACM Trans. Softw. Eng. Methodol.},
month = feb
}

@INPROCEEDINGS{tan2024,
  author={Tan, Zixuan and Zhou, Jiayuan and Hu, Xing and Pan, Shengyi and Liu, Kui and Xia, Xin},
  booktitle={2025 IEEE/ACM 47th International Conference on Software Engineering (ICSE)}, 
  title={Similar but Patched Code Considered Harmful: The Impact of Similar but Patched Code on Recurring Vulnerability Detection and How to Remove Them}, 
  year={2025},
  volume={},
  number={},
  pages={2445-2457},
  doi={10.1109/ICSE55347.2025.00110}}

@inproceedings{primevul,
author = {Ding, Yangruibo and Fu, Yanjun and Ibrahim, Omniyyah and Sitawarin, Chawin and Chen, Xinyun and Alomair, Basel and Wagner, David and Ray, Baishakhi and Chen, Yizheng},
title = {Vulnerability Detection with Code Language Models: How Far Are We?},
year = {2025},
isbn = {9798331505691},
publisher = {IEEE Press},
url = {https://doi.org/10.1109/ICSE55347.2025.00038},
doi = {10.1109/ICSE55347.2025.00038},
booktitle = {Proceedings of the IEEE/ACM 47th International Conference on Software Engineering},
pages = {1729–1741},
numpages = {13},
location = {Ottawa, Ontario, Canada},
series = {ICSE '25}
}

@inproceedings{steenhoek2023empirical,
author = {Steenhoek, Benjamin and Rahman, Md Mahbubur and Jiles, Richard and Le, Wei},
title = {An Empirical Study of Deep Learning Models for Vulnerability Detection},
year = {2023},
isbn = {9781665457019},
publisher = {IEEE Press},
url = {https://doi.org/10.1109/ICSE48619.2023.00188},
doi = {10.1109/ICSE48619.2023.00188},
booktitle = {Proceedings of the 45th International Conference on Software Engineering},
pages = {2237–2248},
numpages = {12},
location = {Melbourne, Victoria, Australia},
series = {ICSE '23}
}

@article{chakraborty2021deep,
  title={Deep learning based vulnerability detection: Are we there yet?},
  author={Chakraborty, Saikat and Krishna, Rahul and Ding, Yangruibo and Ray, Baishakhi},
  journal={IEEE Transactions on Software Engineering},
  volume={48},
  number={9},
  pages={3280--3296},
  year={2021},
  publisher={IEEE}
}

@inproceedings{fu2022linevul,
author = {Fu, Michael and Tantithamthavorn, Chakkrit},
title = {LineVul: a transformer-based line-level vulnerability prediction},
year = {2022},
isbn = {9781450393034},
publisher = {Association for Computing Machinery},
address = {New York, NY, USA},
url = {https://doi.org/10.1145/3524842.3528452},
doi = {10.1145/3524842.3528452},
booktitle = {Proceedings of the 19th International Conference on Mining Software Repositories},
pages = {608–620},
numpages = {13},
location = {Pittsburgh, Pennsylvania},
series = {MSR '22}
}

@inproceedings{feng2020codebert,
    title = "{C}ode{BERT}: A Pre-Trained Model for Programming and Natural Languages",
    author = "Feng, Zhangyin  and
      Guo, Daya  and
      Tang, Duyu  and
      Duan, Nan  and
      Feng, Xiaocheng  and
      Gong, Ming  and
      Shou, Linjun  and
      Qin, Bing  and
      Liu, Ting  and
      Jiang, Daxin  and
      Zhou, Ming",
    editor = "Cohn, Trevor  and
      He, Yulan  and
      Liu, Yang",
    booktitle = "Findings of the Association for Computational Linguistics: EMNLP 2020",
    month = nov,
    year = "2020",
    address = "Online",
    publisher = "Association for Computational Linguistics",
    url = "https://aclanthology.org/2020.findings-emnlp.139/",
    doi = "10.18653/v1/2020.findings-emnlp.139",
    pages = "1536--1547"
}

@inproceedings{guo2022unixcoder,
    title = "{U}ni{X}coder: Unified Cross-Modal Pre-training for Code Representation",
    author = "Guo, Daya  and
      Lu, Shuai  and
      Duan, Nan  and
      Wang, Yanlin  and
      Zhou, Ming  and
      Yin, Jian",
    editor = "Muresan, Smaranda  and
      Nakov, Preslav  and
      Villavicencio, Aline",
    booktitle = "Proceedings of the 60th Annual Meeting of the Association for Computational Linguistics (Volume 1: Long Papers)",
    month = may,
    year = "2022",
    address = "Dublin, Ireland",
    publisher = "Association for Computational Linguistics",
    url = "https://aclanthology.org/2022.acl-long.499/",
    doi = "10.18653/v1/2022.acl-long.499",
    pages = "7212--7225"
}

@inproceedings{hanif2022vulberta,
  author={Hanif, Hazim and Maffeis, Sergio},
  booktitle={2022 International Joint Conference on Neural Networks (IJCNN)}, 
  title={VulBERTa: Simplified Source Code Pre-Training for Vulnerability Detection}, 
  year={2022},
  volume={},
  number={},
  pages={1-8},
  doi={10.1109/IJCNN55064.2022.9892280}}

@inproceedings{steenhoek2024dataflow,
author = {Steenhoek, Benjamin and Gao, Hongyang and Le, Wei},
title = {Dataflow Analysis-Inspired Deep Learning for Efficient Vulnerability Detection},
year = {2024},
isbn = {9798400702174},
publisher = {Association for Computing Machinery},
address = {New York, NY, USA},
url = {https://doi.org/10.1145/3597503.3623345},
doi = {10.1145/3597503.3623345},
booktitle = {Proceedings of the IEEE/ACM 46th International Conference on Software Engineering},
articleno = {16},
numpages = {13},
location = {Lisbon, Portugal},
series = {ICSE '24}
}

@ARTICLE{li2018vuldeepecker,
author={Zou, Deqing and Wang, Sujuan and Xu, Shouhuai and Li, Zhen and Jin, Hai},
journal={ IEEE Transactions on Dependable and Secure Computing },
title={{ VulDeePecker: A Deep Learning-Based System for Multiclass Vulnerability Detection }},
year={2021},
volume={18},
number={05},
ISSN={1941-0018},
pages={2224-2236},
doi={10.1109/TDSC.2019.2942930},
url = {https://doi.ieeecomputersociety.org/10.1109/TDSC.2019.2942930},
publisher={IEEE Computer Society},
address={Los Alamitos, CA, USA},
month=sep}

@article{zou2021interpreting,
author = {Zou, Deqing and Zhu, Yawei and Xu, Shouhuai and Li, Zhen and Jin, Hai and Ye, Hengkai},
title = {Interpreting Deep Learning-based Vulnerability Detector Predictions Based on Heuristic Searching},
year = {2021},
issue_date = {April 2021},
publisher = {Association for Computing Machinery},
address = {New York, NY, USA},
volume = {30},
number = {2},
issn = {1049-331X},
url = {https://doi.org/10.1145/3429444},
doi = {10.1145/3429444},
journal = {ACM Trans. Softw. Eng. Methodol.},
month = mar,
articleno = {23},
numpages = {31}
}

@inproceedings{Ni2023DistinguishingLI,
  title={Distinguishing look-alike innocent and vulnerable code by subtle semantic representation learning and explanation},
  author={Ni, Chao and Yin, Xin and Yang, Kaiwen and Zhao, Dehai and Xing, Zhenchang and Xia, Xin},
  booktitle={Proceedings of the 31st ACM joint european software engineering conference and symposium on the foundations of software engineering},
  pages={1611--1622},
  year={2023}
}

@inproceedings{rahman2024towards,
author = {Rahman, Md Mahbubur and Ceka, Ira and Mao, Chengzhi and Chakraborty, Saikat and Ray, Baishakhi and Le, Wei},
title = {Towards Causal Deep Learning for Vulnerability Detection},
year = {2024},
isbn = {9798400702174},
publisher = {Association for Computing Machinery},
address = {New York, NY, USA},
url = {https://doi.org/10.1145/3597503.3639170},
doi = {10.1145/3597503.3639170},
booktitle = {Proceedings of the IEEE/ACM 46th International Conference on Software Engineering},
articleno = {153},
numpages = {11},
location = {Lisbon, Portugal},
series = {ICSE '24}
}

@inproceedings{risse2024uncovering,
author = {Risse, Niklas and B\"{o}hme, Marcel},
title = {Uncovering the limits of machine learning for automatic vulnerability detection},
year = {2024},
isbn = {978-1-939133-44-1},
publisher = {USENIX Association},
address = {USA},
booktitle = {Proceedings of the 33rd USENIX Conference on Security Symposium},
articleno = {238},
numpages = {18},
location = {Philadelphia, PA, USA},
series = {SEC '24}
}

@article{huang2025survey,
author = {Huang, Lei and Yu, Weijiang and Ma, Weitao and Zhong, Weihong and Feng, Zhangyin and Wang, Haotian and Chen, Qianglong and Peng, Weihua and Feng, Xiaocheng and Qin, Bing and Liu, Ting},
title = {A Survey on Hallucination in Large Language Models: Principles, Taxonomy, Challenges, and Open Questions},
year = {2025},
issue_date = {March 2025},
publisher = {Association for Computing Machinery},
address = {New York, NY, USA},
volume = {43},
number = {2},
issn = {1046-8188},
url = {https://doi.org/10.1145/3703155},
doi = {10.1145/3703155},
journal = {ACM Trans. Inf. Syst.},
month = jan,
articleno = {42},
numpages = {55}
}

@inproceedings{improving,
author = {Du, Yilun and Li, Shuang and Torralba, Antonio and Tenenbaum, Joshua B. and Mordatch, Igor},
title = {Improving factuality and reasoning in language models through multiagent debate},
year = {2024},
publisher = {JMLR.org},
booktitle = {Proceedings of the 41st International Conference on Machine Learning},
articleno = {467},
numpages = {31},
location = {Vienna, Austria},
series = {ICML'24}
}

@inproceedings{liang-etal-2024-encouraging,
    title = "Encouraging Divergent Thinking in Large Language Models through Multi-Agent Debate",
    author = "Liang, Tian  and
      He, Zhiwei  and
      Jiao, Wenxiang  and
      Wang, Xing  and
      Wang, Yan  and
      Wang, Rui  and
      Yang, Yujiu  and
      Shi, Shuming  and
      Tu, Zhaopeng",
    editor = "Al-Onaizan, Yaser  and
      Bansal, Mohit  and
      Chen, Yun-Nung",
    booktitle = "Proceedings of the 2024 Conference on Empirical Methods in Natural Language Processing",
    month = nov,
    year = "2024",
    address = "Miami, Florida, USA",
    publisher = "Association for Computational Linguistics",
    url = "https://aclanthology.org/2024.emnlp-main.992/",
    doi = "10.18653/v1/2024.emnlp-main.992",
    pages = "17889--17904"
}

@inproceedings{wan-etal-2025-mamm,
    title = "MAMM-Refine: A Recipe for Improving Faithfulness in Generation with Multi-Agent Collaboration",
    author = "Wan, David  and
      Chen, Justin  and
      Stengel-Eskin, Elias  and
      Bansal, Mohit",
    editor = "Chiruzzo, Luis  and
      Ritter, Alan  and
      Wang, Lu",
    booktitle = "Proceedings of the 2025 Conference of the Nations of the Americas Chapter of the Association for Computational Linguistics: Human Language Technologies (Volume 1: Long Papers)",
    month = apr,
    year = "2025",
    address = "Albuquerque, New Mexico",
    publisher = "Association for Computational Linguistics",
    url = "https://aclanthology.org/2025.naacl-long.498/",
    doi = "10.18653/v1/2025.naacl-long.498",
    pages = "9882--9901",
    ISBN = "979-8-89176-189-6"
}

@article{huang2023large,
  title={Large language models cannot self-correct reasoning yet},
  author={Huang, Jie and Chen, Xinyun and Mishra, Swaroop and Zheng, Huaixiu Steven and Yu, Adams Wei and Song, Xinying and Zhou, Denny},
  journal={arXiv preprint arXiv:2310.01798},
  year={2023}
}

@article{he2025llm,
author = {He, Junda and Treude, Christoph and Lo, David},
title = {LLM-Based Multi-Agent Systems for Software Engineering: Literature Review, Vision, and the Road Ahead},
year = {2025},
issue_date = {June 2025},
publisher = {Association for Computing Machinery},
address = {New York, NY, USA},
volume = {34},
number = {5},
issn = {1049-331X},
url = {https://doi.org/10.1145/3712003},
doi = {10.1145/3712003},
journal = {ACM Trans. Softw. Eng. Methodol.},
month = may,
articleno = {124},
numpages = {30}
}

@article{wen2024evalsva,
  title={Evalsva: Multi-agent evaluators for next-gen software vulnerability assessment},
  author={Wen, Xin-Cheng and Ye, Jiaxin and Gao, Cuiyun and Wu, Lianwei and Liao, Qing},
  journal={arXiv preprint arXiv:2501.14737},
  year={2024}
}

@INPROCEEDINGS{mao2024multi,
  author={Mao, Zhenyy and Li, Jialong and Jin, Dongming and Li, Munan and Tei, Kenji},
  booktitle={2024 IEEE 24th International Conference on Software Quality, Reliability, and Security Companion (QRS-C)}, 
  title={Multi-Role Consensus Through LLMs Discussions for Vulnerability Detection}, 
  year={2024},
  volume={},
  number={},
  pages={1318-1319},
  doi={10.1109/QRS-C63300.2024.00173}}

@article{wei2024llm,
author={Wei, Zhiyuan and Sun, Jing and Sun, Yuqiang and Liu, Ye and Wu, Daoyuan and Zhang, Zijian and Zhang, Xianhao and Li, Meng and Liu, Yang and Li, Chunmiao and Wan, Mingchao and Dong, Jin and Zhu, Liehuang},
journal={ IEEE Transactions on Software Engineering },
title={{ Advanced Smart Contract Vulnerability Detection via LLM-Powered Multi-Agent Systems }},
year={2025},
volume={51},
number={10},
ISSN={1939-3520},
pages={2830-2846},
doi={10.1109/TSE.2025.3597319},
url = {https://doi.ieeecomputersociety.org/10.1109/TSE.2025.3597319},
publisher={IEEE Computer Society},
address={Los Alamitos, CA, USA},
month=oct
}

@inproceedings{ma2025combining,
author = {Ma, Wei and Wu, Daoyuan and Sun, Yuqiang and Wang, Tianwen and Liu, Shangqing and Zhang, Jian and Xue, Yue and Liu, Yang},
title = {Combining Fine-Tuning and LLM-Based Agents for Intuitive Smart Contract Auditing with Justifications},
year = {2025},
isbn = {9798331505691},
publisher = {IEEE Press},
url = {https://doi.org/10.1109/ICSE55347.2025.00027},
doi = {10.1109/ICSE55347.2025.00027},
booktitle = {Proceedings of the IEEE/ACM 47th International Conference on Software Engineering},
pages = {1742–1754},
numpages = {13},
location = {Ottawa, Ontario, Canada},
series = {ICSE '25}
}

@inproceedings{hu2023large,
  title={Large language model-powered smart contract vulnerability detection: New perspectives},
  author={Hu, Sihao and Huang, Tiansheng and {\.I}lhan, Fatih and Tekin, Selim Furkan and Liu, Ling},
  booktitle={2023 5th IEEE International Conference on Trust, Privacy and Security in Intelligent Systems and Applications (TPS-ISA)},
  pages={297--306},
  year={2023},
  organization={IEEE},
  doi = {10.1109/TPS-ISA58951.2023.00044}
}

@article{widyasari2025let,
  title={Let the trial begin: A mock-court approach to vulnerability detection using llm-based agents},
  author={Widyasari, Ratnadira and Weyssow, Martin and Irsan, Ivana Clairine and Ang, Han Wei and Liauw, Frank and Ouh, Eng Lieh and Shar, Lwin Khin and Kang, Hong Jin and Lo, David},
  journal={arXiv preprint arXiv:2505.10961},
  year={2025},
}

@inproceedings{wang2025causalrag,
    title = "{C}ausal{RAG}: Integrating Causal Graphs into Retrieval-Augmented Generation",
    author = "Wang, Nengbo  and
      Han, Xiaotian  and
      Singh, Jagdip  and
      Ma, Jing  and
      Chaudhary, Vipin",
    editor = "Che, Wanxiang  and
      Nabende, Joyce  and
      Shutova, Ekaterina  and
      Pilehvar, Mohammad Taher",
    booktitle = "Findings of the Association for Computational Linguistics: ACL 2025",
    month = jul,
    year = "2025",
    address = "Vienna, Austria",
    publisher = "Association for Computational Linguistics",
    url = "https://aclanthology.org/2025.findings-acl.1165/",
    doi = "10.18653/v1/2025.findings-acl.1165",
    pages = "22680--22693",
    ISBN = "979-8-89176-256-5"
}

@inproceedings{wan2025digest,
    title = "Digest the Knowledge: Large Language Models empowered Message Passing for Knowledge Graph Question Answering",
    author = "Wan, Junhong  and
      Yu, Tao  and
      Jiang, Kunyu  and
      Fu, Yao  and
      Jiang, Weihao  and
      Zhu, Jiang",
    editor = "Che, Wanxiang  and
      Nabende, Joyce  and
      Shutova, Ekaterina  and
      Pilehvar, Mohammad Taher",
    booktitle = "Proceedings of the 63rd Annual Meeting of the Association for Computational Linguistics (Volume 1: Long Papers)",
    month = jul,
    year = "2025",
    address = "Vienna, Austria",
    publisher = "Association for Computational Linguistics",
    url = "https://aclanthology.org/2025.acl-long.750/",
    doi = "10.18653/v1/2025.acl-long.750",
    pages = "15426--15442",
    ISBN = "979-8-89176-251-0"
}

@inproceedings{lewis2020retrieval,
author = {Lewis, Patrick and Perez, Ethan and Piktus, Aleksandra and Petroni, Fabio and Karpukhin, Vladimir and Goyal, Naman and K\"{u}ttler, Heinrich and Lewis, Mike and Yih, Wen-tau and Rockt\"{a}schel, Tim and Riedel, Sebastian and Kiela, Douwe},
title = {Retrieval-augmented generation for knowledge-intensive NLP tasks},
year = {2020},
isbn = {9781713829546},
publisher = {Curran Associates Inc.},
address = {Red Hook, NY, USA},
booktitle = {Proceedings of the 34th International Conference on Neural Information Processing Systems},
articleno = {793},
numpages = {16},
location = {Vancouver, BC, Canada},
series = {NIPS '20}
}

@article{gao2023retrieval,
  title={Retrieval-augmented generation for large language models: A survey},
  author={Gao, Yunfan and Xiong, Yun and Gao, Xinyu and Jia, Kangxiang and Pan, Jinliu and Bi, Yuxi and Dai, Yi and Sun, Jiawei and Wang, Meng and Wang, Haofen},
  journal={arXiv preprint arXiv:2312.10997},
  year={2023}
}

@inproceedings{ovadia2024fine,
  title={Fine-tuning or retrieval? comparing knowledge injection in llms},
  author={Ovadia, Oded and Brief, Menachem and Mishaeli, Moshik and Elisha, Oren},
  booktitle={Proceedings of the 2024 conference on empirical methods in natural language processing},
  pages={237--250},
  year={2024}
}

@inproceedings{cvefixes,
author = {Bhandari, Guru and Naseer, Amara and Moonen, Leon},
title = {CVEfixes: automated collection of vulnerabilities and their fixes from open-source software},
year = {2021},
isbn = {9781450386807},
publisher = {Association for Computing Machinery},
address = {New York, NY, USA},
url = {https://doi.org/10.1145/3475960.3475985},
doi = {10.1145/3475960.3475985},
booktitle = {Proceedings of the 17th International Conference on Predictive Models and Data Analytics in Software Engineering},
pages = {30–39},
numpages = {10},
location = {Athens, Greece},
series = {PROMISE 2021}
}

@inproceedings{ni2024megavul,
author = {Ni, Chao and Shen, Liyu and Yang, Xiaohu and Zhu, Yan and Wang, Shaohua},
title = {MegaVul: A C/C++ Vulnerability Dataset with Comprehensive Code Representations},
year = {2024},
isbn = {9798400705878},
publisher = {Association for Computing Machinery},
address = {New York, NY, USA},
url = {https://doi.org/10.1145/3643991.3644886},
doi = {10.1145/3643991.3644886},
booktitle = {Proceedings of the 21st International Conference on Mining Software Repositories},
pages = {738–742},
numpages = {5},
location = {Lisbon, Portugal},
series = {MSR '24}
}

@article{li2024cleanvul,
  title={Cleanvul: Automatic function-level vulnerability detection in code commits using llm heuristics},
  author={Li, Yikun and Zhang, Ting and Widyasari, Ratnadira and Tun, Yan Naing and Nguyen, Huu Hung and Bui, Tan and Irsan, Ivana Clairine and Cheng, Yiran and Lan, Xiang and Ang, Han Wei and others},
  journal={arXiv preprint arXiv:2411.17274},
  year={2024}
}

@inproceedings{johnson2013don,
  title={Why don't software developers use static analysis tools to find bugs?},
  author={Johnson, Brittany and Song, Yoonki and Murphy-Hill, Emerson and Bowdidge, Robert},
  booktitle={2013 35th International Conference on Software Engineering (ICSE)},
  pages={672--681},
  year={2013},
  organization={IEEE}
}

@article{goseva2015capability,
author = {Goseva-Popstojanova, Katerina and Perhinschi, Andrei},
title = {On the capability of static code analysis to detect security vulnerabilities},
year = {2015},
issue_date = {December 2015},
publisher = {Butterworth-Heinemann},
address = {USA},
volume = {68},
number = {C},
issn = {0950-5849},
doi = {10.1016/j.infsof.2015.08.002},
journal = {Inf. Softw. Technol.},
month = dec,
pages = {18–33},
numpages = {16}
}

@article{kaur2020comparative,
title = {A Comparative Study of Static Code Analysis tools for Vulnerability Detection in C/C++ and JAVA Source Code},
journal = {Procedia Computer Science},
volume = {171},
pages = {2023-2029},
year = {2020},
note = {Third International Conference on Computing and Network Communications (CoCoNet'19)},
issn = {1877-0509},
doi = {10.1016/j.procs.2020.04.217},
url = {https://www.sciencedirect.com/science/article/pii/S1877050920312023},
author = {Arvinder Kaur and Ruchikaa Nayyar}
}

@article{VKG,
author = {Yin, Jiao and Hong, Wei and Wang, Hua and Cao, Jinli and Miao, Yuan and Zhang, Yanchun},
title = {A Compact Vulnerability Knowledge Graph for Risk Assessment},
year = {2024},
issue_date = {September 2024},
publisher = {Association for Computing Machinery},
address = {New York, NY, USA},
volume = {18},
number = {8},
issn = {1556-4681},
url = {https://doi.org/10.1145/3671005},
doi = {10.1145/3671005},
journal = {ACM Trans. Knowl. Discov. Data},
month = jul,
articleno = {194},
numpages = {17}
}

@inproceedings{CKGD,
    title = "Constructing a Knowledge Graph from Textual Descriptions of Software Vulnerabilities in the National Vulnerability Database",
    author = "H{\o}st, Anders  and
      Lison, Pierre  and
      Moonen, Leon",
    editor = {Alum{\"a}e, Tanel  and
      Fishel, Mark},
    booktitle = "Proceedings of the 24th Nordic Conference on Computational Linguistics (NoDaLiDa)",
    month = may,
    year = "2023",
    address = "T{\'o}rshavn, Faroe Islands",
    publisher = "University of Tartu Library",
    url = "https://aclanthology.org/2023.nodalida-1.40/",
    pages = "386--391",
    doi = {10.48550/arXiv.2305.00382}
}

@inproceedings{CTI-KG,
  author={Rastogi, Nidhi and Dutta, Sharmishtha and Gittens, Alex and Zaki, Mohammed J. and Aggarwal, Charu},
  booktitle={2022 IEEE International Conference on Trust, Security and Privacy in Computing and Communications (TrustCom)}, 
  title={TINKER: A framework for Open source Cyberthreat Intelligence}, 
  year={2022},
  volume={},
  number={},
  pages={1569-1574},
  doi={10.1109/TrustCom56396.2022.00225}}

@article{wu2025vulnsc,
author = {Wu, Bozhi and Liu, Chengjie and Li, Zhiming and Cao, Yushi and Sun, Jun and Lin, Shang-Wei},
title = {Enhancing Vulnerability Detection via Inter-procedural Semantic Completion},
year = {2025},
issue_date = {July 2025},
publisher = {Association for Computing Machinery},
address = {New York, NY, USA},
volume = {2},
number = {ISSTA},
url = {https://doi.org/10.1145/3728912},
doi = {10.1145/3728912},
journal = {Proc. ACM Softw. Eng.},
month = jun,
articleno = {ISSTA037},
numpages = {23}
}

%%
%% If your work has an appendix, this is the place to put it.
% \appendix
% \begin{figure}[t]
%   \centering
%   \includegraphics[width=\columnwidth]{Figure/Appendix/Appendix1.png}
%   \caption{Prompt template for knowledge node extraction.}
%   \label{fig:appendix-knowledge-extraction}
%   \Description{appendix1}
% \end{figure}

% \begin{figure}[t]
%   \centering
%   \includegraphics[width=\columnwidth]{Figure/Appendix/Appendix2.png}
%   \caption{Prompt template for cross-CVE global edge construction.}
%   \label{fig:appendix-global-edge}
%   \Description{appendix2}
% \end{figure}

\end{document}